\documentclass{kapedbk} 

\usepackage{epsf}

\let\footnote\savefootnote

\let\footnotetext\savefootnotetext

\normallatexbib

\begin{document}
\articletitle{Transport in Single Channel~ \\  Quantum Wires}


\author{Hermann Grabert}
\affil{Fakult{\"a}t f{\"u}r Physik\\
Albert--Ludwigs--Universit{\"a}t\\
Hermann--Herder--Stra{\ss}e 3\\
D-79104 Freiburg, Germany}
\email{grabert@physik.uni-freiburg.de}


\begin{abstract}
This tutorial article gives an introduction to the methods needed to
treat interacting electrons in a quantum wire with a single occupied
band. Since one--dimensional Fermions cannot be described in terms of 
noninteracting quasiparticles, the Tomonaga--Luttinger model is
presented in some detail with an emphasis on transport properties. 
To achieve a self--contained presentation, the Bosonization technique 
for one--dimensional Fermions is developed, accentuating features relevant 
for nonequilibrium systems. The screening of an impurity in the wire is
discussed, and the insight gained on the electrostatics of a 
quantum wire is used to describe the coupling to Fermi--liquid
reservoirs. These parts of the article should be readily accessible
to students with a background in quantum mechanics including second
quantization. To illustrate the usefulness of the methods
presented, the current--voltage relation is determined exactly for
a spin--polarized quantum wire with a particular value of the
interaction parameter. This part requires familiarity with path
integral techniques and connects with the current literature.
\end{abstract}


\noindent
{\bf To appear in:} {\it Exotic States in 
Quantum Nanostructures}\, ed.\ by S.~Sarkar, Kluwer, Dordrecht.

\section{Introduction}
Within the last decade there has been increased interest in the
behavior of quasi one--dimensional Fermionic systems, due to
significant advances in the fabrication of single channel
quantum wires \cite{goni,tarucha,yacobi} based on
semiconductor heterostructures
and the observation of non--Fermi liquid behavior in carbon nanotubes
\cite{delft,berkeley,basel}. While the unusual equilibrium properties
of Fermions in one dimension have been studied since many
decades and are well documented in review articles
\cite{solyom,voit}, nonequilibrium quantum wires are an area
of active research with many important question remaining to be answered.

In this article, we give a rather elementary introduction to the
theoretical framework underlying much of the present studies on
transport properties of one--dimensional Fermions. While we do
not review extensively features of the Tomonaga--Luttinger model
\cite{tomonaga,luttinger}
upon which these studies are based, we give an elementary introduction
to the Bosonization technique which is an essential ingredient of
current theoretical methods. We do not review the rather long
history of Bosonization starting with the work by
Schotte and Schotte \cite{schotte} in 1969. Some important articles
are contained in a book of reprints collected by Stone \cite{stone}.
Our approach is based on  Haldane's algebraic Bosonization \cite{haldane},
which can be understood with the usual graduate level background in
physics. For a more in--depth discussion of the method, we
refer to a recent review by von Delft and Schoeller
\cite{delftschoeller}. The field theoretical approach to Bosonization,
which is probably harder to learn but easier to apply, has lately
been expounded by Gogolin, Nersesyan and Tsvelik \cite{gogolin}.

We employ the Bosonization technique to describe a quantum wire
coupled to Fermi--liquid reservoirs. In this connection the
electrostatic properties of the wire play an important role.
Landauer's approach \cite{landauer} to transport in mesoscopic 
systems, which is based on Fermi liquid theory, is generalized to take 
the electronic correlations in a single channel quantum wire into account.
In the Bosonized version of the model the coupling to reservoirs
is shown to be described in terms of radiative boundary
conditions \cite{eggergrabert}. This allows us to use the
powerful Bosonization method also for nonequilibrium wires.
About a decade ago, Kane and Fisher \cite{kanefisher} have noted
that transport properties of one--dimensional Fermions are 
strongly affected by impurities. Even weak impurities have a dramatic
effect at sufficiently low temperatures leading to a zero bias 
anomaly of the conductance. It is the aim of the article to present the 
theoretical background necessary to study the recent literature on 
this subject. Again, we do not provide a review of
transport properties of the Tomonaga--Luttinger model. Rather,
the methods developed are illustrated by treating a particular case.

\subsection{Noninteracting electrons in one dimension}
Let us start by considering first noninteracting 
electrons of mass $m$ moving along a one--dimensional
wire with a scatterer at $x=0$. For simplicity, the scattering
potential is taken as a $\delta$--potential
\begin{equation}
V_{{\rm sc}}(x) = \frac{\hbar^2}{m} \, \Lambda \, \delta (x) \label{vsc}
\end{equation}
where $\Lambda$ characterizes the strength. The Schr\"odinger
equation
\begin{equation}
- \, \frac{\hbar^2}{2m} \, \psi^{\prime \prime} (x) +
\frac{\hbar^2}{m}  \Lambda \, \delta (x) \, \psi (x) = \varepsilon
\, \psi (x)
\end{equation}
has for all positive energies
\begin{equation}
\varepsilon_k = \frac{\hbar^2 k^2}{2m}
\end{equation}
a solution ($k > 0$)
\begin{equation}
\psi_k (x) = \frac{1}{\sqrt{2 \pi}} \ \left\{ \begin{array}{lll}
e^{ikx} \ + r_k e^{-ikx} & , x < 0 &\\
t_k \ e^{ikx} & , x > 0 &
\end{array} \right.
\end{equation}
describing a wave incident from the left that is partially
transmitted and partially reflected. The transmission amplitude
\begin{equation}
t_k = \frac{1}{1 +i \Lambda /k} \ = \sqrt{T_k} \ e^{i \eta_k}
\end{equation}
determines the transmission coefficient $T_k$ and the phase shift
$\eta_k$. Likewise, there is a solution ($k > 0$)
\begin{equation}
\psi_{-k} (x) = \psi_k (-x)
\end{equation}
describing a wave incident from the right.

When the ends of the wire are connected to electrodes with
electrical potentials $\mu_L$ and $\mu_R$, a voltage
\begin{equation}
U = (\mu_L - \mu_R)/e
\end{equation}
is applied to the wire, where $e$ is the electron charge, and an
electrical current
\begin{equation}
I = e \ \int^{+ \infty}_{- \infty} dk \, f_k  j_k
\end{equation}
flows. Here
\begin{equation}
j_k (x) = {\rm Im} \,\psi^*_k(x) \frac{\hbar}{m} \,
\frac{\partial}{\partial x} \, \psi_k (x) = \frac{\hbar k}{2 \pi
m} \, T_k
\end{equation}
is the particle current in state $\psi_k$ which is independent of
$x$, and the
\begin{eqnarray}
f_k & = & \left\{ \begin{array}{lll}
f(\varepsilon_k - \mu_L) & ,  k > 0 &\\
f(\varepsilon_k - \mu_R) & ,  k < 0 &
\end{array} \right.
\end{eqnarray}
with $f (\varepsilon) = 1/ ( e^{\beta \varepsilon}+1)$
are state occupation probabilities determined by the
Fermi function of the electrode from which the particles come.
Both electrodes are assumed to be at the same inverse temperature
$\beta$. Putting $\mu_{L,R} = \varepsilon_F \pm \frac{1}{2}e
U$, we readily find for small voltages $U$
\begin{equation}
I = G\,U
\end{equation}
with the conductance
\begin{eqnarray}
G & = & e^2 \int^{\infty}_0 dk \, \frac{\hbar k}{2 \pi m} \, T_k
\frac{\partial}{\partial \varepsilon_F} \, f(\varepsilon_k -
\varepsilon_F) \nonumber \\
&&\\
 & = & \frac{e^2}{h} \, \int^{\infty}_{0} d \varepsilon \,
T(\varepsilon) \left[-  \frac{\partial}{\partial \varepsilon} \,
f(\varepsilon - \varepsilon_F) \right].
\nonumber
\end{eqnarray}
Provided $\beta \varepsilon_F$ is large, this yields the Landauer
formula for a single transport channel
\begin{equation}
G = \frac{e^2}{h} \, T_F ,
\end{equation}
where $T_F$ is the transmission coefficient at the Fermi energy
$\varepsilon_F$. If we take into account the spin degeneracy of
real electrons, the conductance becomes multiplied by 2.
\subsection{Fano--Anderson Model}
Of course, to describe electrons in one
dimension, we may also start from a tight binding model with
localized electronic states at positions $x_j = aj$ where $a$ is
the lattice constant. The Hamiltonian in the presence of an
impurity at $x=0$ then takes the form of the
Fano-Anderson model \cite{fano,anderson}
\begin{equation}
H = \varepsilon^{}_0\, a^{\dagger}_0a^{}_0 -
t \sum^{+ \infty}_{j=- \infty} \
(a^{\dagger}_{j+1}  a^{}_j + {\rm h.c.}) ,
\end{equation}
where $t$ is the hopping matrix element and $\varepsilon_0 > 0$
is the extra energy needed to occupy the perturbed site at $x=0$.
The $a_j$  are Fermi operators obeying the usual anti--commutation
relations. This model is also exactly solvable \cite{mahan} with
eigenstates in the energy band
\begin{equation}
\varepsilon_k = -2t \cos (ka) \ ,  |k| \le \frac{\pi}{a},
\end{equation}
and the transmission amplitude takes the form
\begin{equation}
t_k = \frac{1}{1+i/ \lambda_k} \, ,
\end{equation}
where
\begin{equation}
\lambda_k = \frac{2t}{\varepsilon_0} \, \sin (ka) = \frac{\hbar
v_k}{\varepsilon_0a}.
\end{equation}
The last relation follows by virtue of $v_k = (1/ \hbar) \partial
\varepsilon_k/ \partial k$. It is now easily checked that this
model leads to the same result for the linear conductance than
the free electron model examined previously, provided we match
parameters of the unperturbed models such that the Fermi
velocities $v_F$ at the two Fermi points coincide, and we adjust
the strength of the $\delta$--function in Eq.~(\ref{vsc}) such
that
\begin{equation}
\frac{\hbar^2}{m} \, \Lambda = \varepsilon_0 a.
\end{equation}
Then, we end up with the same transition coefficient $T_F$ at the
Fermi energy.
\subsection{Noninteracting Tomonaga--Luttinger model}
\begin{figure}
\hspace{1.5cm}
\epsfysize=5cm
\epsffile{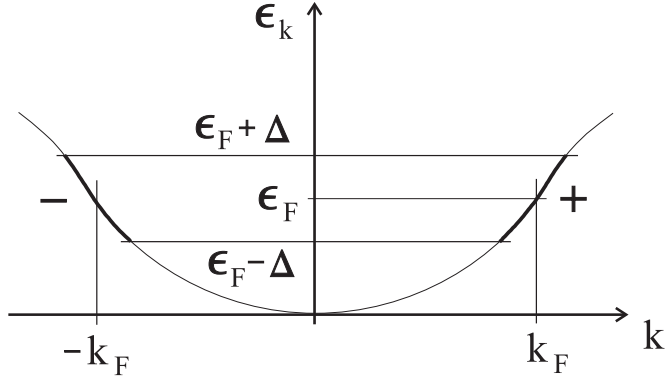}
\caption[Branches]{Physical energy dispersion curve $\epsilon_k$ (thin line) and the two branches $+,-$ (thick lines) of the TL model}
\label{branches}
\end{figure}

Apparently, for low temperatures and small applied voltages, the
conductance only depends on properties of states in the vicinity
of the Fermi energy $\varepsilon_F$. We can take advantage of
this fact by introducing still another model, the noninteracting
Tomonaga--Luttinger (TL) model, which has the same properties near
the two Fermi points $\pm k_F$ but is more convenient once we
introduce electron interactions. Let us formally decompose the
true energy dispersion curve $\varepsilon_k$ into two branches
$+$ and $-$ of right--moving and left--moving electrons,
respectively, where these branches comprise  states in the energy
interval $\left[ \varepsilon_F - \Delta  ,  \varepsilon_F +
\Delta \right]$. (cf.\ Fig.~\ref{branches}).
If $\Delta$ is chosen
large enough, these branches should  suffice to describe the
low energy
physics of the true physical model, since for low temperatures
and small applied voltages states with energy below
$\varepsilon_F - \Delta$ are always occupied while states above
$\varepsilon_F + \Delta$ are empty.
The Hamiltonian of the noninteracting TL model reads
\begin{equation}
H = \sum_p \sum_{k} \, \varepsilon_{p,k}
\left[ c_{p,k}^{\dagger}c_{p,k}^{} - \langle c^{\dagger}_{p,k}c_{p,k}^{}
\rangle_0 \right]  .
\label{tlnon}
\end{equation}
Here $p = \pm$ labels the two branches. We have introduced a
quantization length $L$ such that wave vectors are 
discrete\footnote{We use periodic boundary conditions here and consider
the limit of large $L$ in the sequel. The same techniques can
also be used for other boundary conditions 
\cite{delftschoeller,fabriziogogolin}.}
\begin{equation}
k = \frac{2 \pi}{L} \, n_k \ , \ n_k \ \mbox{integer.}
\end{equation}
Further, the $\varepsilon_{\pm}(k)$ are single particle energies
measured relative to the Fermi energy $\varepsilon_F$,  i.e.,
$\varepsilon_+(k_F) = \varepsilon_-(-k_F)=0$. The  operators
$c^{\dagger}_{p,k}$ and $c_{p,k}$ are Fermi creation and
annihilation operators obeying anti--commutation
relations, in particular
\begin{equation}
\left[ c_{p,k}^{}\, , \, c^{\dagger}_{p^{\prime}\!,k^{\prime}}
\right]_+ = \delta_{p,p^{\prime}} \delta_{k,k^{\prime}} .
\end{equation}
The sum over $k$ states in Eq.~(\ref{tlnon}) is restricted to $k$
values near $\pm k_F$ such that $\varepsilon_{\pm}(k) \in [-
\Delta \, , \, \Delta]$. Finally, the ground state energy is
subtracted in Eq.~(\ref{tlnon}) where the ground state $|0, 0
\rangle_0$ is defined by
\begin{eqnarray}
&& c_{+,k}|0,0 \rangle_0 =
c_{-,-k} |0,0 \rangle_0 = 0 \ \
\mbox{for} \ k >
k_F \, , \nonumber \\
&&\\
 && c^{\dagger}_{+,k}|0,0 \rangle_0 = c^{\dagger}_{-,-k}|0,0 \rangle_0 =
0 \ \ \mbox{for} \ k \le k_F \, . \nonumber
\end{eqnarray} Frequently,
the spectra $\varepsilon_{\pm,k}$ are linearized about the Fermi
points, i.e.,
\begin{equation}
\varepsilon_{\pm,k} = \pm \hbar v_F(k \mp k_F) \ ,
\label{linspec}
\end{equation}
and then the cutoff energy $\Delta$ is used as a large energy
scale regularizing divergent expressions. Since
for low energy systems only inert empty or occupied states are
added, an increase of $\Delta$ is admissible. However,
the linearization of the spectra is only realistic
in the close vicinity of the Fermi points. In fact, some phenomena
not discussed here, e.g.\ the thermopower
\cite{fisherkane}, depend on band curvature. We shall come back to
the limitations of the linearization (\ref{linspec}) below.

Before we discuss transport properties of the TL model, we first
introduce methods, the advantage of which becomes apparent only
when we pass on to the case of interacting electrons. These
methods are independent of the precise dispersion law as long as the
$\varepsilon_{p,k}$ are  monotonous functions of $k$.

\section{Bosonization}

The noninteracting TL model allows for a formulation in terms 
of Bose operators. 
We discuss the Bosonization technique here for spinless
Fermions first, and then extend it to the spinful case.
\subsection{Density operators and their algebra}
\begin{figure}
\hspace{1.2cm}
\epsfysize=5cm
\epsffile{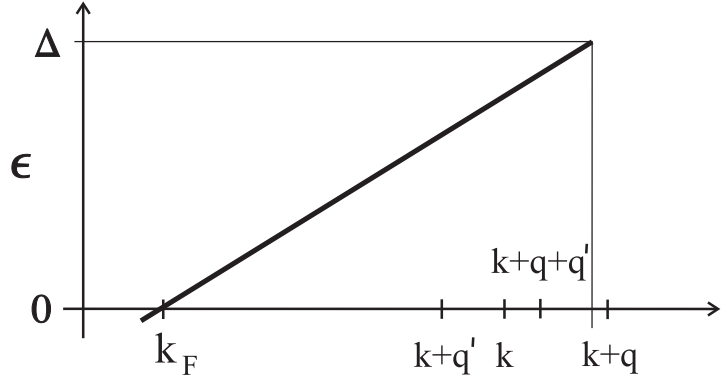}
\caption{Wave vectors in the sum (\ref{Fzwei}) for $q > 0$ and
$q^{\prime} < 0$ near the upper cutoff energy of the $+$ branch.}
\label{rand}
\end{figure}

While the physical problem of one--dimensional Fermions is
described by a single energy dispersion curve $\varepsilon_k$
with empty states at both ends of the range of $k$ values, the TL
model introduces two branches with empty states at one end but
occupied states at the other end of the $k$ range. This leads to
unusual algebraic properties we will discuss now. Since the range
of allowed $k$ values plays an important role in this discussion,
we keep track of it in detail by introducing
\begin{equation}
W_{p,k} = \left\{ \begin{array}{l} 1 \ \ \mbox{for} \
\varepsilon_{p,k} \in [- \Delta \, , \, \Delta] \\
0 \ \ \mbox{else} \end{array} \right. . \label{wcutoff}
\end{equation}
Let us define Fourier components of the densities of $p$--movers
$(p= \pm)$ by
\begin{equation}
\tilde{\rho}_{p,q} = \sum_k \ W_{p,k}W_{p,k+q}\,
c^{\dagger}_{p,k}c^{}_{p,k+q} \, . \label{rotilde}
\end{equation}
Since Fermi operators on different branches anti--commute, we have
\begin{equation}
\left[ \tilde{\rho}_{p,q} \, , \,
\tilde{\rho}_{p^{\prime}\!,q^{\prime}} \right]_-=0 \ \ \mbox{for} \
p \neq p^{\prime}\, .
\end{equation}
On the other hand, for the commutator  on the same, say the $+$
branch, we find
\begin{eqnarray}
 \left[ \tilde{\rho}_{+,q} , \tilde{\rho}_{+,q^{\prime}}
\right]_- &=& \sum_{kk^{\prime}} \,
W_{+,k}W_{+,k+q}W_{+,k^{\prime}}W_{+,k^{\prime}+q^{\prime}}
 \\
&\times& \Big[c^{\dagger}_{+,k} c_{+,k+q}^{}
c^{\dagger}_{+,k^{\prime}} c^{}_{k^{\prime}+q^{\prime}} -
c^{\dagger}_{+,k^{\prime}} c^{}_{+,k^{\prime}+q^{\prime}}
c^{\dagger}_{+,k} c^{}_{+,k+q} \Big]\, \nonumber .
\end{eqnarray}
We now use anti--commutation relations for the second and third
operator in each of the two products of four Fermi operators,
e.g., $c_{+,k+q}^{}c^{\dagger}_{+,k^{\prime}} = \delta_{k+q  ,
k^{\prime}}-c^{\dagger}_{+,k^{\prime}}c^{}_{+,k+q}$. Then, the
remaining products of four Fermi operators are easily seen to
cancel by virtue of the anti--commutation relations, and the two
terms with two Fermi operators can be written as
\begin{eqnarray}
&&\left[ \tilde{\rho}_{+,q} \, , \, \tilde{\rho}_{+,q^{\prime}}
\right]_- \\
&&= \sum_k \, W_{+,k}W_{+,k+q+q^{\prime}}  \left[
W_{+,k+q}^2 - W_{+,k+q^{\prime}}^2 \right]  c^{\dagger}_{+,k}
c^{}_{+,k+q+q^{\prime}} \, . \nonumber \label{Fzwei}
\end{eqnarray}
The terms of the sum are finite only for $W_{+,k} =
W_{+,k+q+q^{\prime}} = 1$. But then $W_{+,k+q}$ and
$W_{+,k+q^{\prime}}$ are also equal to $1$ if $q$ and
$q^{\prime}$ have the same sign. In this case all terms of the
sum (\ref{Fzwei}) vanish. Hence, a nontrivial commutator can only
arise if $q$ and $q^{\prime}$ have different signs, say $q > 0$
and $q^{\prime} < 0$. Then non--vanishing terms of the sum
(\ref{Fzwei}) may occur near the upper cutoff where
$\varepsilon_{+,k} \approx \Delta$, in particular, when
\begin{displaymath}
W_{+,k} = W_{+,k+q+q^{\prime}} = W_{+,k+q^{\prime}} = 1\, , \
\ \mbox{but} \ W_{+,k+q}=0 ,
\end{displaymath}
as illustrated in Fig.~\ref{rand}. However, in this case the
operator $c^{\dagger}_{+,k}c^{}_{+,k+q+q^{\prime}}$ tries to
annihilate a particle in the state $k+q+q^{\prime}$ with an
energy near $\Delta$. In the low energy sector of the model these
states are always empty and $c_{+,k+q+q^{\prime}}$ can be
replaced by zero. Hence, for a low energy system we get no
contribution to the commutator from states near the upper cutoff
energy.

A contribution near the lower cutoff energy may arise from terms
with $W_{+,k}=W_{+,k+q+q^{\prime}}=W_{+,k+q}=1$ but $W_{+,k+q^{\prime}}=0$
since $\varepsilon_{+,k+q^{\prime}} < - \Delta$. In this case we
use the anti--commutation relations to write Eq.~(\ref{Fzwei}) as
\begin{eqnarray}
 \left[ \tilde{\rho}_{+,q} \, , \, \tilde{\rho}_{+,q^{\prime}}
\right]_-
& =& \sum_k \, W_{+,k}W_{+,k+q+q^{\prime}}  \left[
W_{+,k+q}^2-W_{+,k+q^{\prime}}^2 \right]  \nonumber \\
&\times& \left[ \delta_{q,-q^{\prime}} -
c^{}_{+,k+q+q^{\prime}}c^{\dagger}_{+,k} \right] \, . \label{commlow}
\end{eqnarray}
With a similar argument as above, the operator $c^{\dagger}_{+,k}$
may now be replaced by zero since near the lower cutoff all
states are occupied. We thus obtain
\begin{equation}
 \left[ \tilde{\rho}_{+,q} \, , \, \tilde{\rho}_{+,q^{\prime}}
\right]_-
=
 \delta_{q , -q^{\prime}}  \sum_{k<k_F} \, W_{+,k}^2
\left[W_{+,k+q}^2-W_{+,k-q}^2 \right] \, ,
\end{equation}
where the condition $k<k_F$ ensures that we do not get
contributions from the region near the upper cutoff where the
Fermi operators in Eq.~(\ref{commlow}) cannot be dropped.
Finally, since
\begin{equation}
\sum_{k<k_F} \ W_{+,k}^2 W_{+,k+q}^2 - \sum_{k<k_F} \ W_{+,k-q}^2
W_{+,k}^2 = n_q \, ,
\end{equation}
where $q = \frac{2 \pi}{L} \, n_q$ determines the number $n_q$ of
additional non--vanishing terms in the first sum, and with a
similar reasoning for the $-$ branch and other signs of $q$, we
get
\begin{equation}
\left[ \tilde{\rho}_{p,q}\, , \,
\tilde{\rho}_{p^{\prime}\!
,q^{\prime}} \right]_- = p \, \delta_{p,p^{\prime}} \delta_{q  , -
q^{\prime}} \, n_q  . \label{commro}
\end{equation}
As we have seen this nontrivial commutator only arises since we
have introduced two branches with empty states at one end and
occupied states at the other end. The result (\ref{commro}) can
also be derived in the same way when the sharp cutoff functions
$W_{\pm,k}$ defined in Eq.~(\ref{wcutoff}) are replaced by a
smooth cutoff. Furthermore, the relation (\ref{commro}) is not an
exact operator relation but holds only in the low energy sector
of the Fock space. However, for technical convenience, we may
use the linearized spectrum (\ref{linspec}) and send the cutoff
$\Delta$ to infinity. We then obtain a model where the relations
(\ref{commro}) hold as formally exact commutation relations but
have to remember that only the low energy properties of this
model are related to the physical problem of one--dimensional
electrons.

\subsection{Bose operators and basis vectors}
\begin{figure}
\hspace{2.5cm}
\epsfysize=5cm
\epsffile{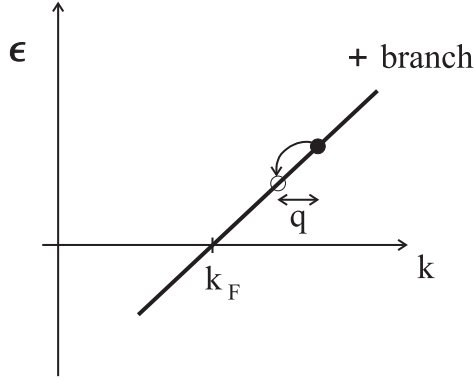}
\caption{Action of $b_q$ for $q > 0$ on states of the $+$
branch.}
\label{bbild}
\end{figure}

\begin{figure}
\hspace{1.8cm}
\epsfysize=7cm
\epsffile{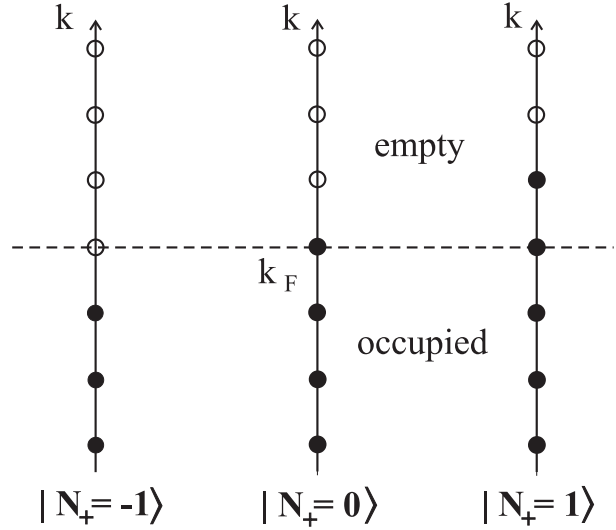}
\caption{Graphical representation of $N_+$--particle ground
states of right--movers.}
\label{nstates}
\end{figure}

With the result (\ref{commro}) it is now straightforward to
introduce Bose annihilation and creation operators\footnote{As usual, 
$\theta(x)$ denotes the step function: 
$\theta(x)=1$ for $x > 0$, $\theta(x)=0$ for $x<0$.}
\begin{eqnarray}
b_q &=& \frac{-i}{\sqrt{|n_q|}} \, \sum_p \, \theta (pq)
\tilde{\rho}_{p,q} \, , \nonumber \\
&&\\
 b^{\dagger}_q &=& \frac{i}{\sqrt{|n_q|}} \, \sum_p \, \theta (pq)
\tilde{\rho}_{p,-q}\, , \nonumber \label{bvonq}
\end{eqnarray}
where $q \neq 0$. 
Inserting (\ref{rotilde}), we see that for $q>0$ the operator
$b_{q}$ lowers the wave vector $k$ of right--movers by $q$,
provided the state $k-q$ is empty, cf.\ Fig.~\ref{bbild}, while
$b_{-q}$ acts accordingly on left--movers.  By virtue of
Eq.~(\ref{commro}), the $b_{q}$ and $b^{\dagger}_q$ satisfy Bose
commutation relations, in particular
\begin{equation}
[ b^{}_q \, , \,  b^{\dagger}_{q^{\prime}}]_- = \delta_{q,q^{\prime}} .
\label{commb}
\end{equation}
When acting on a $N$--particle state, the operators
$b^{\dagger}_q$ and $b_q$ create and annihilate electron--hole
pairs, respectively, but they conserve of course the particle
numbers  $N_p$ of each branch. Consider now $N_+$--particle
ground states of right--movers where $|N_+= 0 \rangle_0$ is the
many--body state with all single particle states $k \le k_F$
occupied and all states $k > k_F$ empty, while in the state
$|N_+= 1 \rangle_0$ also the level $k=k_F + \frac{2 \pi}{L}$ is
occupied, and so on, as illustrated in Fig.~\ref{nstates}. One can
then show that the states
\begin{equation}
|N_+ , \{ m_{q}\}_{q>0} \rangle = \prod_{q>0} \ \frac{\left(
b^{\dagger}_q \right)^{m_{q}}}{\sqrt{m_{q}\, !}} \ |N_+ \rangle_0 \ ,
\label{nmstate}
\end{equation}
with $N_+ = 0, \ \pm 1, \ \pm 2, \ \ldots$ and $m_{q}=0,1,2, \
\ldots$ form a complete basis in the Fock space of right--movers.
It is obvious that the states (\ref{nmstate}) are all
within the Fock space spanned by the standard basis vectors $| \{
N_{+,k} \} \rangle$, where the $N_{+,k} = 0,1$ are the usual Fermi
occupation numbers, that is eigenvalues of
$c^{\dagger}_{+,k} c^{}_{+,k}$. Haldane \cite{haldane} has demonstrated
completeness of the basis (\ref{nmstate}) with the help of the 
model Hamiltonian
\begin{equation}
h_+ = \sum_k \, (n_k- \frac{1}{2})  : c^{\dagger}_{+,k} c^{}_{+,k} :\ ,
\end{equation}
where
\begin{equation}
:  c^{\dagger}_{+,k} c^{}_{+,k}  : = \left\{ \begin{array}{lll}
c^{\dagger}_{+,k} c^{}_{+,k} & \ \mbox{for} & k > 0 \\
c^{}_{+,k} c^{\dagger}_{+,k} & \ \mbox{for} & k \le 0 \ .
\end{array} \right.
\end{equation}
This Hamiltonian assigns positive energies to all states and allows for an
explicit evaluation of the partition function $Z= \sum_{\alpha} \
\exp (- \beta E_{\alpha})$ for the two sets $\{\alpha \}$ of
basis vectors. Since $Z$ is a sum of positive terms, and the sum
over the states (\ref{nmstate}) gives the same result as the sum
over the states $| \{ N_{+,k} \} \rangle$, the basis $|N_{+}, \{
m_{q} \}_{q>0} \rangle$ spans the entire Fock space of right--movers.

Likewise we may introduce a basis
\begin{equation}
|N_-, \{m_{q}\}_{q<0} \rangle = \prod_{q< 0} \frac{\left(
b^{\dagger}_q\right)^{m_{q}}}{\sqrt{m_{q}\, !}} |N_-\rangle_0
\end{equation} in the Fock space of
left--movers and combine both to the basis
\begin{equation}
|N_+,N_-,\{ m_q \}\rangle =  \prod_{q\ne 0}
\frac{\left(b^{\dagger}_q \right)^{m_q}}{\sqrt{m_q !}}
 \, |N_+,N_- \rangle_0
\label{mbasis}
\end{equation}
in the Fock space of one--dimensional particles.
\subsection{Ladder and particle number operators}
\begin{figure}
\hspace{2.5cm}
\epsfysize=7cm
\epsffile{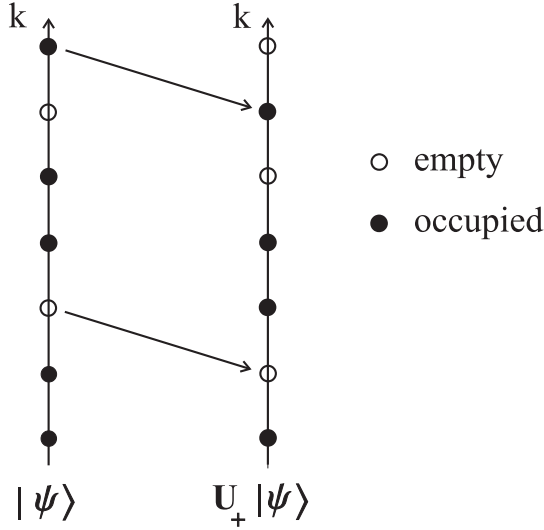}
\caption{Illustration of the action of $U_+$
on a many--body state $| \psi \rangle$ of right--movers.}
\label{uwirk}
\end{figure}

 It is clear that
the Bose operators $b^{}_q$, $b^{\dagger}_q$ cannot generate the whole
algebra of operators in the Fock space since they
preserve the particle numbers $N_p$. Hence, the Bose operators
need to be supplemented by ladder operators removing or adding
a particle of branch $p$. The lowering operators $U_p$ are defined by
\begin{equation}
U_p | N_p,N_{-p} \rangle_0 = p^{N_+}\, | N_p -1, N_{-p} \rangle_0 \ ,
\label{ladder}
\end{equation}
where the sign factor $p^{N_+}$ is one possible choice
assuring anti--commutation relations for Fermi operators on
different branches, and
\begin{equation}
[  U_p\, , \, b^{}_q]_- = [U_p \, ,\, b^{\dagger}_q]_- = 0\, .
\end{equation}
The adjoint raising operator obeys $U_p^{\dagger}=U_p^{-1}$. The action of
the ladder operator $U_+$ on a state in the Fock space of
right--movers is illustrated in Fig.~\ref{uwirk}. Since the ladder
operators commute with the Bose operators $b^{}_q, b^{\dagger}_q$,
they preserve the electron--hole pair excitations present in a state.

Now, the particle number $N_p$ is an eigenvalue of the particle operator
\begin{equation}
\hat{N}_p = \sum_k \, \left( \ c^{\dagger}_{p,k} c^{}_{p,k} - \langle
c^{\dagger}_{p,k} c^{}_{p,k} \rangle_0 \right)\ ,
\end{equation}
where the average $\langle \ \ \rangle_0$ is over the $| N_p = 0
\rangle_0$ ground state. The ladder operators and the particle
numbers satisfy the  commutation relations
\begin{eqnarray}
\left[\hat{N}_p \, , \, U_{p^{\prime}} \right]_- & = & -
\delta_{p,p^{\prime}} U_p \ , \nonumber \\
\left[ \hat{N}_{p}\, , \,  U^{\dagger}_{p^{\prime}} \right]_- & = &
\delta_{p,p^{\prime}} U^{\dagger}_p \ ,
\end{eqnarray}
that are easily shown by applying them to an arbitrary vector from the
basis set (\ref{mbasis}). Furthermore, the ladder operators obey
\begin{eqnarray}
\left[ \ U^{}_+, U^{}_- \right]_+ & =
& \left[ U^{\dagger}_+, U^{\dagger}_- \right]_+ =
0 , \nonumber \\
\left[ \ U^{}_p, U^{\dagger}_{p^{\prime}} \right]_+ & = & 2\,
\delta_{p,p^{\prime}} ,
\label{commu}
\end{eqnarray}
which may be demonstrated in the same way taking the sign factor
in Eq.~(\ref{ladder}) into account.
\subsection{Bosonic phase fields}
With the Bose and ladder
operators we should be able to represent all operators in the
Fock space of the TL model. To demonstrate this one usually
constructs explicitly the Fermi annihilation operators
\begin{equation}
\Psi_p(x) = \frac{1}{\sqrt{L}}  \sum_k \, e^{ikx} \, c_{p,k} .
\label{psi}
\end{equation}
It is convenient to first introduce Bosonic phase 
fields
\begin{equation}
\varphi_p(x) = \frac{1}{2 \pi}  \sum_{q \neq 0} \, \theta (pq)
\frac{1}{\sqrt{|n_q|}} \, e^{iqx} \, b_q ,
\label{kleinphi}
\end{equation}
and associated Hermitian phase fields
\begin{equation}
\Phi_p(x) = \varphi_p(x) + \varphi^{\dagger}_p(x) .
\label{grossphi}
\end{equation}
These are related to the densities
\begin{equation}
\rho_p(x) = \Psi^{\dagger}_p(x) \Psi^{}_p(x) - \langle
\Psi^{\dagger}_p(x) \Psi^{}_p(x) \rangle_0 = \frac{1}{L}  \left[ \
\sum_{q \neq 0} \, e^{iqx} \, \tilde{\rho}_{p,q} + \hat{N}_p
\right]\, , \label{pdensity}
\end{equation}
where the $\tilde{\rho}_{p,q} = \sum_k  c^{\dagger}_{p,k}c^{}_{p,k+q}$
are the Fourier components studied above.  Using
Eq.~(\ref{bvonq}), the phase field (\ref{kleinphi}) can
also be written as
\begin{equation}
\varphi_p(x) = - \frac{i}{2 \pi} \sum_{q \neq 0} \, \theta (pq)
\, \frac{1}{|n_q|} \, e^{iqx} \, \tilde{\rho}_{p,q}\, .
\end{equation}
Because of the factor $\theta (pq)$, we may replace $|n_q|$ by $p
n_q = \frac{L}{2 \pi} pq$. Then, adding the Hermitian conjugate field, 
we readily find for the gradient
\begin{equation}
\frac{\partial}{\partial x} \Phi_p(x)  = \frac{p}{L}  \sum_{q
\neq 0} \, e^{iqx} \, \tilde{\rho}_{p,q}
 =  p \left( \rho_p(x) - \frac{1}{L}  \hat{N}_p \right) .
\label{gradphi}
\end{equation}
The Bosonic phase fields $\varphi (x)$, $\varphi^{\dagger} (x)$
obey the commutation relations
\begin{equation}
\left[ \varphi_p(x) \, , \, \varphi_{p^{\prime}}(x^{\prime})
\right]_- = \left[  \varphi^{\dagger}_p(x) \, , \,
\varphi^{\dagger}_{p^{\prime}}(x^{\prime}) \right]_- = 0 \label{commphia}
\end{equation}
while by virtue of Eqs.~(\ref{commb}) and (\ref{kleinphi}) the
commutator
\begin{equation}
\left[  \varphi_p(x) \, , \,
\varphi^{\dagger}_{p^{\prime}}(x^{\prime}) \right]_-
 =  \frac{\delta_{p,p^{\prime}}}{4 \pi^2}
\sum^{\infty}_{n=1} \, \frac{1}{n} \, e^{ip \frac{2 \pi}{L}
(x-x^{\prime})n} .
\label{commphi}
\end{equation}
Since
\begin{equation}
\sum^{\infty}_{n=1} \, \frac{y^n}{n} \ = - \ln (1-y) \ ,
\end{equation}
the right hand side of Eq.~(\ref{commphi}) is logarithmically
divergent for $x = x^{\prime}$. However, we have to
remember that the sum over $n$ originates from a sum over wave
vectors $q=(2 \pi / L)n$ which due to the energy cutoff are
restricted to small $q$. The effect of a cutoff is seen when we
multiply the terms of the sum (\ref{commphi}) by an exponential
cutoff function $\exp (-aq) = \exp (- \frac{2 \pi}{L} \, a n)$
which limits the sum to wave vectors $q 
 {{\lower 2pt \hbox{$<$} \atop \raise 1pt \hbox{$ \sim$} }} 1/a$, 
where $a$ is a
cutoff length related to a cutoff energy $\Delta \approx \hbar
v_F/a$. We then obtain
\begin{eqnarray}
\left[ \varphi_p(x) \, , \,
\varphi^{\dagger}_{p^{\prime}}(x^{\prime}) \right]_- &=&
- \frac{\delta_{p,p^{\prime}}}{4 \pi^2}  \ln \left( 1 - \exp \left\{ -
\frac{2 \pi}{L} \left[ a - ip (x-x^{\prime}) \right] \right\}
\right) \nonumber \\
&=& - \frac{\delta_{p,p^{\prime}}}{4 \pi^2} \ln\left(\frac{2\pi}{L}
\left[a-ip(x-x^{\prime})\right]\right)
\label{commphib}
\end{eqnarray}
where the second equality holds for $L \gg a$, $|x - x^{\prime}|$.
For later convenience we note that in particular the commutator
for vanishing distance
\begin{equation}
\left[\varphi_p(x) \, ,\, \varphi^{\dagger}_{p^{\prime}}(x) \right]_-
= - \frac{ \delta_{p,p^{\prime}}}{4 \pi^2} \, \ln \frac{2 \pi
a}{L} \label{commphic}
\end{equation}
is cutoff and size dependent.

Simpler algebraic properties are found
for the Hermitian phase field (\ref{grossphi}). From Eqs.~(\ref{commphia})
and (\ref{commphib}) we readily find
\begin{equation}
\left[\Phi_p(x) \, ,\, \Phi_{p^{\prime}}(x^{\prime}) \right]_- = -
  \frac{\delta_{p,p^{\prime}}}{4 \pi^2} \ \ln
\frac{a-ip(x-x^{\prime})}{a+ip(x-x^{\prime})}\ , \label{cutcomm}
\end{equation}
where $L$ has dropped out, and we have for $|x-x^{\prime}| \gg a$
\begin{equation}
\left[ \Phi_p(x) \, ,\, \Phi_{p^{\prime}}(x^{\prime}) \right]_- =
\frac{ip \, \delta_{p,p^{\prime}}}{4 \pi} \, {\rm sign}
(x-x^{\prime}) \, .
\end{equation}
Here  ${\rm sign}(x-x^{\prime})$ is the sign of $x-x^{\prime}$ which
produces a step at $x= x^{\prime}$ that in the presence of a
cutoff is smeared over a length of order $a$.

Instead of the fields $\Phi_{\pm}(x)$ we shall mainly use the
linear combinations
\begin{eqnarray}
\phi(x) & = & \sqrt{\pi}  \left[ \Phi_+(x) + \Phi_-(x) \right]
\nonumber \\
&& \label{phithet} \\
\vartheta (x) & = & \sqrt{\pi}  \left[ \Phi_+(x) - \Phi_-(x)
\right] \nonumber
\end{eqnarray}
that are readily seen to obey the commutation relations
\begin{equation}
\left[ \phi (x) \, , \, \phi (x^{\prime}) \right]_- = \left[
\vartheta(x) \, ,\, \vartheta(x^{\prime}) \right]_- = 0 \nonumber
\end{equation}
and
\begin{equation}
\left[ \phi(x) \, ,\, \vartheta (x^{\prime}) \right]_- = \frac{i}{2}
\, {\rm sign}(x-x^{\prime})\, , \label{commphithet}
\end{equation}
which shows that
\begin{equation}
\Pi_{\vartheta}(x) = - \hbar \frac{\partial}{\partial x} \, \phi
(x) \nonumber
\end{equation}
is the field conjugate to $\vartheta (x)$ with the
canonical commutator
\begin{equation}
\left[\vartheta (x)\, ,\, \Pi_{\vartheta}(x^{\prime}) \right] = i
\hbar \delta (x - x^{\prime}) \, . \nonumber \label{commthetpi}
\end{equation}
Likewise $\Pi_{\phi}(x) = - \hbar \frac{\partial}{\partial x}
\vartheta (x)$ is the field canonically conjugate to $\phi (x)$.
Finally, we note that by virtue of Eq.~(\ref{gradphi})
\begin{equation}
\frac{1}{\sqrt{\pi}} \, \frac{\partial}{\partial x} \, \vartheta
(x) =  \rho_+ (x) + \rho_- (x) - \frac{\hat{N}_++
\hat{N}_-}{L}  \label{gradtheta}
\end{equation}
describes the  fluctuations of the total density
of right-- and left--movers.

\subsection{Bose representation of Fermi operators}

To construct the Fermi operators
$\Psi_p(x)$ in terms of the Bosonic fields, we start by
evaluating the commutator
\begin{equation}
\left[ b_q \, , \, \Psi_p(x) \right]_- = \alpha_{p,q}(x) \,
\Psi_p(x) \, ,\label{commbpsi}
\end{equation}
where
\begin{equation}
\alpha_{p,  q}(x) = \frac{i}{\sqrt{|n_q|}} \, \theta (pq) \,
e^{- iqx} . \label{alpha}
\end{equation}
This follows readily by inserting Eqs.~(\ref{bvonq}) and
(\ref{psi}) into (\ref{commbpsi}) with the help of the Fermi
commutator $\left[ \, \tilde{\rho}_{p^{\prime}\! ,q}, \, c_{p, k}
\right]_- = - \delta_{p,p^{\prime}} \, c_{p,k+q}$. In the same way
we find
\begin{equation}
\left[ b^{\dagger}_q \, , \, \Psi_p (x) \right]_- = \alpha^*_{p,
q} (x) \, \Psi_p (x) \, . \label{commkreuz}
\end{equation}
Operating with $b_q \Psi_p (x)$ onto a $N$--particle ground state
$|N_+,  N_- \rangle_0$ we have
\begin{eqnarray}
&& b_q \Psi_p (x) | N_+, N_- \rangle_0 = \left[ b_q\, , \,
\Psi_p (x) \right]_- | N_+,  N_- \rangle_0 \nonumber \\
&& \label{bpsi} \\
&& \qquad = \alpha_{p, q} (x) \Psi_p (x) | N_+, N_- \rangle_0 ,
\nonumber
\end{eqnarray}
where the first relation holds because $b_q | N_+ ,  N_-
\rangle_0 = 0$. Now, Eq.~(\ref{bpsi}) shows that $\Psi_p (x) |
N_+,  N_- \rangle_0$ is an eigenstate of $b_q$ with eigenvalue
$\alpha_{p,  q} (x)$. Eigenstates of Bose annihilation operators
are known as coherent states, and from their familiar properties
it follows that $\Psi_p (x) | N_+,  N_- \rangle_0$ is of the form
\begin{equation}
\Psi_p(x) | N_+,  N_- \rangle = \lambda_p (x) \exp \left(
\sum_{q \neq 0} \, \alpha_{p,  q} (x) \, b^{\dagger}_q \right)
| N_p - 1,  N_{-p} \rangle_0 \, , \label{psian}
\end{equation}
where $\lambda_p (x)$ is an as yet undetermined normalization
factor, and we have taken into account that $\Psi_p(x)$ reduces
the number $N_p$ of $p$--movers by $1$. With the help of the Bose
commutator (\ref{commb}), which implies
\begin{equation}
\left[b_q \, , \, \exp  \left( \sum_{q^{\prime} \neq 0}
\alpha_{p,q^{\prime}} (x) \, b^{\dagger}_{q^{\prime}} \right)
\right]_- = \alpha_{p,  q} (x) \, \exp \left(
\sum_{q^{\prime} \neq 0} \alpha_{p,  q^{\prime}} (x) \,
b^{\dagger}_{q^{\prime}} \right) \ , \nonumber
\end{equation}
it is readily seen that the ansatz~(\ref{psian}) indeed satisfies
Eq.~(\ref{bpsi}). From the definition (\ref{kleinphi}) of the
Bosonic phase field $\varphi_p (x)$ and Eq.~(\ref{alpha}) we find
\begin{equation}
\sum_{q \neq 0} \, \alpha_{p, q} (x)\, b^{\dagger}_q = 2 \pi i
\, \varphi^{\dagger}_p (x)\, , \nonumber
\end{equation}
and hence from Eq.~(\ref{psian})
\begin{equation}
_0\langle N_p -1, N_{-p} |\, \Psi_p (x)\, | N_p, N_{-p} \rangle_0
= \lambda_p (x)\, , \label{lambda}
\end{equation}
where we have made use of
\begin{equation}
_0\langle N_p -1, N_{-p} |\, \exp  \left[ 2 \pi i \,
\varphi^{\dagger}_p (x) \right] \, | N_p -1,  N_{-p} \rangle_0 =
1 , \nonumber
\end{equation}
which is easily seen by expanding the exponential. We now insert
the representation (\ref{psi}) of $\Psi_p (x)$ into
Eq.~(\ref{lambda}) and use
\begin{equation}
_0\langle N_p -1, N_{-p} | \, c_{p, k}\, | N_p, N_{-p}
\rangle_0 = p^{N_+} \, \delta_{k,  k_p} \, ,\nonumber
\end{equation}
where
\begin{equation}
k_p = p \left( \, k_F + \frac{2 \pi}{L} \, N_p \right) .
\nonumber
\end{equation}
The sign factor $p^{N_+}$ arises from the
anti--commutation relations of Fermi operators on different
branches in accordance with Eq.~(\ref{ladder}). This yields
\begin{equation}
\lambda_p (x) = \frac{1}{\sqrt{L}} \, p^{N_+} \, e^{ip\left(k_F+
\frac{2 \pi}{L} \, N_p\right)x}  , \nonumber
\end{equation}
which can now be combined with the definition (\ref{ladder}) of the
lowering operator to obtain from Eq.~(\ref{psian})
\begin{equation}
\Psi_p(x) | N_{+},  N_- \rangle_0 = \frac{1}{\sqrt{L}} \,
e^{ip\left(k_F+\frac{2\pi}{L}\, N_p\right)x +
2\pi i \varphi^{\dagger}_p(x)} U_p |N_+,N_-\rangle_0 .
\label{psinull}
\end{equation}
This represents the action of a Fermi annihilation operator on a
$N$--particle ground state in terms of Bose and ladder operators.
Since particle--hole excitations are created by Bose operators,
it is straightforward to generalize Eq.~(\ref{psinull}) for
arbitrary states in the Fock space. For a basis vector
(\ref{mbasis}) we have as a consequence of the commutator
(\ref{commkreuz})
\begin{eqnarray}
&& \Psi_p (x) | N_{+}, N_-, \{m_q \} \rangle \nonumber \\
&& \qquad = \sum_{q \neq 0} \, \frac{\left[ \, b^{\dagger}_q -
\alpha^*_{p, q}(x) \right]^{m_q}}{\sqrt{m_q\, !}} \, \Psi_p(x) |
N_+, N_- \rangle_0 \, . 
\end{eqnarray}
Using Eq.~(\ref{psinull}), we find
\begin{eqnarray}
&& \Psi_p(x) | N_+, N_-,  \{ m_q \} \rangle \\
&& = \frac{1}{\sqrt{L}} \, U_p \, e^{ip(k^{}_F+ \frac{2
\pi}{L} \, N^{}_p)x + 2 \pi i \varphi_p^{\dagger}(x)}
\sum_{q \neq 0} \, \frac{\left[ \,
b^{\dagger}_q - \alpha^*_{p, q}(x) \right]^{m_q}}{\sqrt{m_q\, !}}
\, | N_+, N_- \rangle_0 \label{psieins} ,\nonumber
\end{eqnarray}
since the Bose creation operators commute with $\varphi^{\dagger}_p(x)$
and $U_p$. In view of
\begin{equation}
\sum^{}_{q \neq 0} \, \alpha^*_{p, q}(x) b^{}_q = - 2 \pi i \,
\varphi^{}_p(x) 
\end{equation}
and the familiar Bose relation $e^{\alpha b} \,
b^{\dagger} \, e^{- \alpha b} = b^{\dagger} + \alpha$ we have
\begin{equation}
e^{2 \pi i \varphi_p(x)} \, b_q^{\dagger} \, e^{-2 \pi i
\varphi_p(x)} = b^{\dagger}_q - \alpha^{*}_{p, q}(x) \, ,
{\nonumber}
\end{equation}
and Eq.~(\ref{psieins}) may be written as
\begin{eqnarray}
&& \Psi_p(x) | N_+, N_-, \{ m_q \} \rangle \\
&&  = \frac{1}{\sqrt{L}} \, U_p \, e^{ip(k_F+ \frac{2
\pi}{L} \, \hat{N}_p)x} \, e^{2 \pi i \varphi^{\dagger}_p(x)} \,
e^{2 \pi i \varphi_p(x)} \, | N_+,  N_-,  \{ m_q \} \rangle .
\nonumber
\label{psizwei}
\end{eqnarray}
Here we have made use of $\varphi_p(x) | N_+,N_- \rangle_0 = 0$,
and have replaced in the exponent $N_p$ by the operator $\hat{N}_p$,
which is appropriate since $\hat{N}_p$ commutes with the
Bose operators and thus acts on an eigenstate with eigenvalue
$N_p$. Now, the operator acting on the basis vector on the
right hand side
of Eq.~(\ref{psizwei}) has the same form for any vector, and we thus
obtain the operator identity
\begin{equation}
\Psi_p(x) = \frac{1}{\sqrt{L}} \, U_p \, e^{ip(k^{}_F+ \frac{2
\pi}{L} \, \hat{N}^{}_p)x} \, e^{2 \pi i \varphi^{\dagger}_p(x)}
\, e^{2 \pi i \varphi^{}_p(x)} . \label{psidrei}
\end{equation}
This is the desired representation of the Fermi operators
$\Psi_p(x)$ in terms of Bose and ladder operators. The result
(\ref{psidrei}) is in normal ordered form with all Bose
creation operators to the left of the annihilation operators. It
should be noted that this remarkable relation is independent of
the form of the Hamiltonian, it just relates the form
(\ref{psi}) of the Fermi operator with an obvious meaning in
the standard occupation number basis $| \{ N_{+, k} \}, \{
N_{-, k}\} \rangle$ of the Fock space to a representation with
an obvious interpretation in the basis $|N_+,  N_-, \{ m_q
\} \rangle$.

For much of the following discussions another form
of $\Psi_p(x)$ is often more convenient. Using
the commutator (\ref{commphic}) and the operator
relation
\begin{equation}
e^A \, e^B = e^{A+B} \, e^{\frac{1}{2} \, [A,B]_-}\, ,
\label{disent}
\end{equation}
which holds if $[A,B]_-$ commutes with $A$ and $B$,
we may transform the relation (\ref{psidrei}) to read
\begin{equation}
\Psi_p(x) = \frac{1}{\sqrt{2 \pi a}} \, U_p \, e^{ip(k^{}_F +
\frac{2 \pi}{L} \, \hat{N}^{}_p)x + 2 \pi i \Phi^{}_p (x)}\ ,
\nonumber
\end{equation}
where $\Phi_p(x)$ was introduced in Eq.~(\ref{grossphi}).
Finally, in terms of the phase fields $\phi (x)$ and $\vartheta
(x)$ defined in Eq.~(\ref{phithet}), we have
\begin{equation}
\Psi_p(x) = \frac{1}{\sqrt{2 \pi a}} \, U_p \, e^{ip \left[
k^{}_F x + \frac{2 \pi}{L} \, \hat{N}^{}_p x + \sqrt{\pi}
\vartheta (x) \right] +i \sqrt{\pi} \phi (x)} , \nonumber
\label{psiboson}
\end{equation}
which is the form employed mostly in the
literature.\footnote{Some authors use non--standard definitions
of the Fermi annihilation operator fields (\ref{psi}). To 
compare with our notation, 
one has to make proper replacements, e.g., replace $x$ by $-x$.
In addition, the fields $\vartheta(x)$ and $\phi(x)$ are
sometimes defined the other way round.}

\subsection{Bose representation of the Hamiltonian}

In the following, we shall make use explicitly of the Hamiltonian
(\ref{tlnon}) with the linearized energy dispersion curves
(\ref{linspec}). From Eq.~({\ref{bvonq}) we see that the Bose
creation operators may be written as
\begin{equation}
b^{\dagger}_q = \frac{i}{\sqrt{|n_q|}} \, \sum_{pk}  \theta
(pq) \, c^{\dagger}_{p,  k+q} \, c^{}_{p,  k} . \nonumber
\end{equation}
With the help of the Fermi anti--commutation relations we then
find
\begin{equation}
\left[ H_0 \, , \, b^{\dagger}_q \right]_- = \hbar v_F \, |q|
\, b^{\dagger}_q , \label{commhb}
\end{equation}
where we have also made use of $\varepsilon_{p, k+q} -
\varepsilon_{p, k} = p \hbar v_F q$ which holds for the linear
spectrum (\ref{linspec}). Now, a $N$--particle ground state
$|N_+, N_- \rangle_0$ is an eigenstate of $H_0$ with the energy
\begin{equation}
E_0(N_+,  N_-) =  \hbar v_F \, \frac{\pi}{L}
\sum_p \, N_p(N_p+1) \, . \label{enull}
\end{equation}
Since $|0, 0 \rangle_0$ has zero energy by definition and the
single particle energies (\ref{linspec}) are counted relative to
the Fermi energy, the result (\ref{enull}) is obtained readily
by adding for $N_p > 0$ the single particle energies of the
states occupied additionally in the $N$--particle ground state
$|N_+, \, N_- \rangle_0$, while for $N_p < 0$ we have to subtract
the (negative) energies of the particles removed. Using the
commutator (\ref{commhb}) we obtain for an arbitrary basis
vector (\ref{mbasis})
\begin{eqnarray}
&&H_0  \prod^{}_{q \neq 0}  \frac{\left(  b^{\dagger}_q
\right)^{m_q}}{\sqrt{m_q\, !}} \, | N_+,  N_- \rangle_0
\nonumber \\
&&\ \  = 
\prod^{}_{q \neq 0}  \frac{\left( b^{\dagger}_q
\right)^{m_q}}{\sqrt{m_q \, !}} \, \left( H_0 + \sum_{q \neq
0} \, \hbar v_F |q| m_q \right)  \left| N_+,
N_- \rangle_0 \right. ,
\end{eqnarray}
which shows that $|N_+, \, N_-, \, \{ m_q \} \rangle$ is an
eigenstate of $H_0$ with eigenvalue
\begin{equation}
E_0 (N_+, N_-, \{ m_q \}) = E_0 (N_+, N_-) + \hbar
v_F \sum_{q \neq 0} \ |q|m_q  . \label{enullm}
\end{equation}
Since the basis (\ref{mbasis}) is complete and the basis
vectors are clearly eigenstates of
\begin{equation}
H_0 = \hbar v_F \left[ \, \sum_{q \neq 0} \, |q| \,
b^{\dagger}_qb^{}_q \, + \, \frac{\pi}{L}  \sum_p \,
\hat{N}_p(\hat{N}_p+1) \right] \label{tlbose}
\end{equation}
with the proper eigenvalues (\ref{enullm}), we see that
Eq.~(\ref{tlbose}) gives indeed a Bose representation of the
Hamiltonian (\ref{tlnon}).

Combining Eqs.~(\ref{bvonq}) and
(\ref{gradphi}), we may write the gradient of the Bosonic phase
field $\Phi_p(x)$ as
\begin{equation}
\frac{\partial}{\partial x} \, \Phi_p(x) = \frac{p}{\sqrt{2 \pi
L}} \, \sum_{q \neq 0} \, \theta (pq)  \sqrt{|q|}  \left[
i \, e^{iqx} \, b_q + \, {\rm h.c.} \right] , \nonumber
\end{equation}
which gives
\begin{equation}
\sum_p \, \int\limits^{L/2}_{-L/2} dx \left( \frac{\partial
\Phi_p(x)}{\partial x} \right)^2 = \frac{1}{2 \pi}  \sum_{q
\neq 0} \, |q|  \left(
b^{}_qb^{\dagger}_q+b^{\dagger}_qb^{}_q \right) \, .
\end{equation}
The Hamiltonian (\ref{tlbose}) may thus be written as
\begin{equation}
H_0 = \pi \hbar v_F  \int\limits^{L/2}_{-L/2}  dx \, \sum_p  \left[
\, :  \left( \frac{\partial \Phi_p}{\partial x} \right)^2
 : \, + \, \frac{1}{L^2} \, \hat{N}_p (\hat{N}_p+1)
\right]\ , \nonumber
\end{equation}
where $: \ :$ puts the Bose operators in normal order. Further,
in terms of the fields (\ref{phithet}) this reads
\begin{equation}
H_0 = \frac{\hbar v_F}{2} \int\limits^{L/2}_{-L/2}  dx  \left[ :
\left(  \frac{\partial \phi}{\partial x} \right)^2 \, + \,
\left(  \frac{\partial \vartheta}{\partial x} \right)^2  : \,
+ \, \frac{2 \pi}{L^2} \sum_p \hat{N}_p  (\hat{N}_p+1) \right]\, .
\label{hamnull}
\end{equation}
Since $\Pi_{\vartheta} = - \hbar \partial\phi/\partial x$ is
the conjugate density to the phase field $\vartheta$ with
the canonical commutator (\ref{commthetpi}), we finally
obtain the $\vartheta$--representation of the Hamiltonian
\begin{equation}
H_0 = \frac{\hbar v_F}{2}  \int\limits^{L/2}_{-L/2}  dx  \left[ :
\frac{1}{\hbar^2}\Pi_{\vartheta}^2 \, + \,
\left(  \frac{\partial \vartheta}{\partial x} \right)^2  : \,
+ \, \frac{2 \pi}{L^2} \sum_p \hat{N}_p  (\hat{N}_p+1) \right]\, .
\end{equation}
Likewise, using $\Pi_{\phi} = - \hbar \partial\vartheta/\partial x$,
we can readily write down a $\phi$--re\-presentation of $H_0$.

\subsection{Action functional}

In the usual way, we may introduce a Lagrangian
\begin{equation}
L_0 = \int\limits^{L/2}_{-L/2}  dx  \, \Pi_{\vartheta}
\frac{\partial \vartheta}{\partial t} \ - H_0 ,
\nonumber
\end{equation}
where the time rate of change of the $\vartheta$--field reads
\begin{equation}
\frac{\partial}{\partial t} \, \vartheta = \frac{i}{\hbar} \,
\left[  H_0\, , \, \vartheta \right]_- = - v_F \,
\frac{\partial}{\partial x} \, \phi = \frac{v_F}{\hbar} \,
\Pi_{\vartheta}\, , \nonumber
\end{equation}
which follows from the commutation relations (\ref{commphithet}).
In the limit $L \rightarrow \infty$ this gives
\begin{equation}
L_0 = \hbar  \int dx \left[ \frac{1}{2v_F}
\left( \frac{\partial \vartheta}{\partial t} \right)^2 \, - \,
\frac{v_F}{2} \left( \frac{\partial \vartheta}{\partial x} \right)^2
\, -\, \pi v_F \left(
\overline{\rho}^2_+ + \overline{\rho}^2_- \right) \right] \nonumber
\end{equation}
where
\begin{equation}
\overline{\rho}_p = \frac{N_p}{L} \label{rhop}
\end{equation}
is the average density of $p$--movers. Note that by definition the
densities $\overline{\rho}_{p}$ vanish in the ground state
$|0,0 \rangle_0$, where the Fermi levels are at $\pm k_F$. The
$\overline{\rho}_p$ determine a shift of the Fermi points, while
the density fluctuations $\rho_p(x) - \overline{\rho}_p$ 
arising from electron--hole pair excitations are described by
the phase field $\vartheta$. However, we may define a shifted
phase field
\begin{equation}
\vartheta^{\prime} = \vartheta + \sqrt{\pi} \left(
\overline{\rho}_+\, +\, \overline{\rho}_- \right) x \,  - \, \sqrt{\pi}
\left( \overline{\rho}_+\, - \, \overline{\rho}_- \right) v_Ft
\label{thetashift}
\end{equation}
with the properties
\begin{eqnarray}
\frac{1}{\sqrt{\pi}} \, \frac{\partial
\vartheta^{\prime}}{\partial x}  & = & \rho_+ + \rho_-
\nonumber
\\
&& \label{ablthet} \\
\frac{1}{\sqrt{\pi}} \, \frac{\partial
\vartheta^{\prime}}{\partial t}  & = & - v_F  \left(
\rho_+ - \rho_- \right) \, . \nonumber
\end{eqnarray}
Now, the gradient determines the total density of right-- and
left--movers including the ground state density $\overline{\rho}_++
\overline{\rho}_-$, while the time rate of change is proportional
to the particle current. The Lagrangian then takes the simple form
\begin{equation}
L_0 = \frac{\hbar}{2}  \int dx  \left[ \frac{1}{v_F}
\left(  \frac{\partial \vartheta}{\partial t} \right)^2 - v_F
\left(  \frac{\partial \vartheta}{\partial x} \right)^2 \right]\ ,
\nonumber
\end{equation}
where we have omitted the prime. The noninteracting TL model with
dispersionless spectrum (\ref{linspec}) can thus be
characterized by the classical action functional
\begin{equation}
S_0 = \frac{\hbar}{2} \int dt \int dx \, \left[  \frac{1}{v_F} \left(
 \frac{\partial \vartheta}{\partial t} \right)^2 - \,
v_F  \left(  \frac{\partial \vartheta}{\partial x}
\right)^2 \right]\, , \label{action}
\end{equation}
which is the action of a harmonic string with wave velocity $v_F$
and a dimensionless displacement field $\vartheta (x,t)$ measured
in units of $(\hbar/v_F \mu)^{1/2}$ where $\mu$ is the mass
density in the string. From this mechanical analogue it is
obvious that Bosonization provides an alternative description of
one--dimensional Fermions in terms of charge density oscillations
rather than electron--hole pair excitations.

We remark that in the general case of an arbitrary
dispersion law $\varepsilon_{p,k}$ the Bosonization
identities, in particular the representation (\ref{psiboson}) of
the Fermi operators remain valid, however, the Hamiltonian $H_0$
is no longer quadratic in the Bose operators $b_q$,
$b_q^{\dagger}$. As a consequence, the mechanical analogue will
be an anharmonic string which may, of course, be treated in the
harmonic approximation when we restrict ourselves to low energy
excitations. In combination with the Feynman path integral
representation the classical action functional (\ref{action})
can be a convenient starting point for quantum mechanical calculations.

\subsection{Electron density operator}

When we use the TL model to make
predictions for one--dimensional fermions, we have to remember
that the model provides a local approximation to the physical
model in the vicinity of the Fermi points, while there is only a
single energy dispersion curve for real electrons, cf.\
Fig.~\ref{branches}. This has consequences for observables like the
density operator
\begin{equation}
\rho(x) = \frac{1}{L}  \sum_{k,k^{\prime}} \,
e^{-i(k-k^{\prime})x} \left(  a^{\dagger}_ka^{}_{k^{\prime}} -
\langle a^{\dagger}_ka^{}_{k^{\prime}} \rangle_0 \right)\ ,
\label{density}
\end{equation}
where the $a^{}_k$, $a^{\dagger}_{k^{\prime}}$ are Fermi operators
for the underlying physical model with a single branch, and we
have subtracted the constant density of the ground state with the
Fermi points at $\pm k_F$. When we restrict ourselves to low
energy states, the operators $a^{\dagger}_k a^{}_{k^{\prime}}$ can give a
nonvanishing contribution only if both wave vectors $k$ and
$k^{\prime}$ are in the vicinity of the two Fermi points $\pm
k_F$. However, for $k \approx p k_F$, $k^{\prime} \approx
p^{\prime} k_F$, $(p$, $p^{\prime}\! = \pm)$, the operators
$a^{\dagger}_k$, $a^{}_{k^{\prime}}$ may be replaced by the Fermi
operators $c^{\dagger}_{p,k}$, $c^{}_{p^{\prime}\! ,k^{\prime}}$ of the
TL model, and Eq. (\ref{density}) splits into four terms
\begin{eqnarray}
\rho (x) & = & \sum_{pp^{\prime}} \, \frac{1}{L}
\sum_{kk^{\prime}} \, e^{-i(k-k^{\prime})x}  \left(
c^{\dagger}_{p,k}  c^{}_{p^{\prime}\! ,k^{\prime}} \, - \, \langle
c^{\dagger}_{p,k}  c^{}_{p^{\prime}\! ,k^{\prime}} \rangle_0 \right)
\nonumber \\
&& \nonumber \\
& = & \sum_{pp^{\prime}}  \left( \Psi^{\dagger}_p(x)
\Psi^{}_{p^{\prime}}(x) \, - \, \langle \Psi^{\dagger}_p(x)
\Psi^{}_{p^{\prime}}(x) \rangle_0 \right)\, , \label{densityreal}
\end{eqnarray}
where we have used Eq.~(\ref{psi}) to obtain the second line. Now,
the diagonal terms $(p=p^{\prime})$ just give the densities
(\ref{pdensity}) of $p$--movers, while for the non--diagonal terms
we employ the representation (\ref{psiboson}) to find
\begin{equation}
\Psi^{\dagger}_p(x) \Psi^{}_{-p}(x)  =  \frac{1}{2 \pi a} \, e^{-2
i p k_Fx} \, e^{-2ip \sqrt{\pi} \vartheta (x)} \,
 e^{-ip \frac{2 \pi}{L} \, ( \hat{N}_p+
\hat{N}_{-p}+1)} \, U^{\dagger}_pU^{}_{-p} \, ,\nonumber
\end{equation}
where we have used the commutators (\ref{commu}) and the fact
that $\vartheta (x)$ and $\phi (x)$ commute at the same position
$x$, which is seen from Eq.~(\ref{commphithet}). Note that in the
presence of a cutoff the sign function in Eq.~(\ref{commphithet})
takes the form of the right hand side of Eq.~(\ref{cutcomm}) and
therefore vanishes for vanishing argument. The
density (\ref{densityreal}) may thus be written as
\begin{equation}
\rho (x) = \rho_+(x) + \rho_-(x) + \rho_{2k_F}(x)\, , \label{densitye}
\end{equation}
where
\begin{equation}
\rho_{2k_F}(x) = \frac{1}{2 \pi a}  \left(  e^{-2i \left[
k_Fx+ \sqrt{\pi} \vartheta (x) + \frac{\pi}{L} \,  \left(
\hat{N}_p+ \hat{N}_{-p} + 1 \right) \right]} \,
U^{\dagger}_+U^{}_{-} + {\rm h.c.} \right)\, . \label{rhoka}
\end{equation}
Hence, the density operator for real electrons is not just the sum
of the densities in the two branches of the TL model, but there
is an additional $2k_F$-component $\rho_{2k_F}(x)$ which comes
from the fact that right-- and left--movers are propagating in the same
channel and interfere.

In the limit $L \rightarrow \infty$ with
constant average densities (\ref{rhop}), we may introduce the
shifted phase field (\ref{thetashift}) with $t=0$ since the
expression (\ref{densitye}) gives the operator in the
Schr\"odinger picture. We then have in view of Eq.~(\ref{ablthet})
\begin{equation}
\rho (x) = \frac{1}{\sqrt{\pi}} \, \frac{\partial}{\partial x} \,
\vartheta (x) + \rho_{2k_F}(x) \, , \label{rhoboson}
\end{equation}
where we have again suppressed the prime on
$\vartheta$ which now contains the terms in the exponent of
Eq.~(\ref{rhoka}) that depend on the particle numbers.
The operators $U^{\dagger}_pU^{}_{-p}$ in Eq.~(\ref{rhoka})
associated with the scattering of an electron from the $(-p)$--
into the $p$--branch can often be suppressed since changes of the
particle numbers $N_p$ by 1 can be neglected for $L \rightarrow
\infty$. Then $\rho_{2k_F}(x)$ takes the simple form
\begin{equation}
\rho_{2k_F}(x) = \frac{k_F}{\pi} \, \cos \left[ 2k_F x + 2
\sqrt{\pi} \vartheta (x) \right] \, , \label{rhozwei}
\end{equation}
where we have chosen a cutoff length $a=k^{-1}_F$ of order
a typical microscopic length.

\subsection{Fermions with spin}

We now briefly summarize the modifications necessary to include
the electron spin. Then, the electron spectrum has two branches
$s = \uparrow,\downarrow$ for spin up and spin down
particles. For each species $s$, we may proceed exactly as for
spinless fermions and define Bosonic phase fields
$\vartheta_{s}(x)$ and $\phi_{s}(x)$. There are now four
branches of the TL model and correspondingly four particle number
operators $\hat{N}_{p,s}$. When defining the ladder
operators, we have to include appropriate sign factors in the
generalization of Eq.~(\ref{ladder}) to assure anti-commutation
relations
\begin{eqnarray}
\left[ U_{p,s}\, , \, U_{p^{\prime}\! ,s^{\prime}}
\right]_+ & = & 2\, \delta_{p, p^{\prime}} \, \delta_{s,
s^{\prime}} \, \left(U_{p,s}\right)^2 \nonumber \\
&& \nonumber \\
\left[U^{\dagger}_{p,s}\, , \, U^{\dagger}_{p^{\prime}\! ,
s^{\prime}} \right]_+ & = & 2\, \delta_{p,p^{\prime}} \,
\delta_{s,s^{\prime}} \, \left(U^{\dagger}_{p,s}\right)^2 
\label{ladderspin} \\
&& \nonumber \\
\left[ U_{p,s}\, , \, U^{\dagger}_{p^{\prime}\! ,
s^{\prime}} \right]_+ & = & 2 \, \delta_{p,  p^{\prime}} \,
\delta_{s,s^{\prime}} \, \nonumber
\end{eqnarray}
that extend the relations (\ref{commu}) to the case of four
branches. The Bose representation (\ref{psiboson}) of the Fermi
operators takes again the same form for each species $s$, i.e.,
\begin{equation}
\Psi_{p,s}(x) = \frac{1}{\sqrt{2 \pi a}} \, U_{p,s}
\, e^{ip\left[ k_Fx + \frac{2 \pi}{L} \, \hat{N}_{p,s}
\, x + \sqrt{\pi} \vartheta_{s}(x) \right] +i \sqrt{\pi}
\phi_{s}(x)}\, . \label{psispin}
\end{equation}
It is often convenient to transform to the phase fields
\begin{eqnarray}
\vartheta_{\rho}(x) & = & \frac{1}{\sqrt{2}} \left[
\vartheta_{\uparrow}(x)+ \vartheta_{\downarrow}(x) \right] \nonumber \\
&&\label{phaserho}  \\
\phi_{\rho}(x) & = & \frac{1}{\sqrt{2}}  \left[
\phi_{\uparrow}(x) + \phi_{\downarrow}(x) \right] \nonumber
\end{eqnarray}
and
\begin{eqnarray}
\vartheta_{\sigma}(x) & = & \frac{1}{\sqrt{2}} \left[
\vartheta_{\uparrow}(x)- \vartheta_{\downarrow}(x) \right] \nonumber \\
&  & \label{phasesigma} \\
\phi_{\sigma}(x) & = & \frac{1}{\sqrt{2}}  \left[
\phi_{\uparrow}(x) - \phi_{\downarrow}(x) \right] \nonumber
\end{eqnarray}
that satisfy the commutation relations
\begin{eqnarray}
\left[\vartheta_{\rho}(x)\, , \,  \vartheta_{\rho}(x^{\prime}
 ) \right]_- &=& \left[ \vartheta_{\sigma}(x)\, , \,
\vartheta_{\sigma}(x^{\prime}) \right]_- = 0
\nonumber \\
&&\nonumber \\
  \left[
\phi_{\rho}(x)\, ,\, \phi_{\rho}(x^{\prime} ) \right]_- &=& \left[
\phi_{\sigma}(x)\, , \, \phi_{\sigma}(x^{\prime} ) \right]_- =
0 \nonumber \\
&&  \\
 \left[ \vartheta_{\rho}(x)\, , \,
\vartheta_{\sigma}(x^{\prime} ) \right]_- & =& \left[
\phi_{\rho}(x)\, , \, \phi_{\sigma}(x^{\prime} ) \right]_- = 0
\nonumber \\
&& \nonumber \\
 \left[ \vartheta_{\rho}(x)\, , \, \phi_{\sigma}(x^{\prime}  )
\right]_- &=& 0 \nonumber
\end{eqnarray}
and
\begin{equation}
\left[ \vartheta_{\rho}(x)\, , \, \phi_{\rho}(x^{\prime} )
\right] = \left[  \vartheta_{\sigma}(x)\, , \,
\phi_{\sigma}(x^{\prime} ) \right] = \frac{i}{2} \ {\rm sign}
 (x-x^{\prime} ) \, .\nonumber
\end{equation}
These fields describe charge and spin density excitations. In
particular, instead of Eq.~(\ref{gradtheta}) we now have
\begin{eqnarray}
\sqrt{\frac{2}{\pi}} \, \frac{\partial}{\partial x} \,
\vartheta_{\rho}(x)& =& \rho_{+, \uparrow}(x)\,  +\,  \rho_{+,
\downarrow}(x)\, +\, \rho_{-, \uparrow}(x)\, +\, \rho_{-,
\downarrow}(x) \nonumber \\
&& - \frac{\hat{N}_{+, \uparrow}+
 \hat{N}_{+, \downarrow}
+ \hat{N}_{-, \uparrow} + \hat{N}_{-, \downarrow}}{L}\ ,
\end{eqnarray}
which gives the fluctuations of the total particle density of
right-- and left--movers, while
\begin{eqnarray}
\sqrt{\frac{2}{\pi}} \, \frac{\partial}{\partial x} \,
\vartheta_{\sigma}(x)& =& \rho_{+, \uparrow}(x) \, + \, \rho_{-,
 \uparrow}(x)\, -\, \rho_{+, \downarrow}(x)\, -\, \rho_{-, \,
\downarrow}(x) \nonumber \\
&& -  \frac{\hat{N}_{+, \uparrow} + \hat{N}_{-, \uparrow} -
\hat{N}_{+, \downarrow} - \hat{N}_{-, \downarrow}}{L}
\end{eqnarray}
determines fluctuations of the spin density. In the limit $L
\rightarrow \infty$ with given average densities
\begin{equation}
\overline{\rho}_{p,s} = \frac{N_{p,s}}{L} \nonumber
\end{equation}
we may again introduce shifted phase fields
$\vartheta^{\prime}_{s}$ for each species $s$ according to
Eq.~(\ref{thetashift}) and then obtain the classical action
functional of the noninteracting model with dispersionless energy spectrum
as
\begin{equation}
S_0 = \frac{\hbar}{2}  \sum_{\alpha}  \int dt  \int dx  \left[
\frac{1}{v_F}  \left(  \frac{\partial
\vartheta_{\alpha}}{\partial t} \right)^2 - \, v_F
\left( \frac{\partial \vartheta_{\alpha}}{\partial x}
\right)^2 \right]\ , \label{actionspin}
\end{equation}
where the sum is over the two spin directions $\alpha =
\uparrow, \downarrow$. Then, $S_0$ is the sum of the actions
(\ref{action}) for each species $s$. The expression
(\ref{actionspin}) remains, however, also valid if the sum is
over $\alpha = \rho,  \sigma$ with the phase fields introduced
in Eqs.~(\ref{phaserho}) and (\ref{phasesigma}). Hence, the
action functional for spinful electrons can be split in a charge
and spin contribution
\begin{equation}
S_0 = S_{0,  \rho} + S_{0, \sigma}\, , \nonumber
\end{equation}
where each term has the form (\ref{action}) of the action of a
harmonic string.

The representation of the true electron density
operator (\ref{density}) in terms of the Bosonic phase fields has
also a straightforward extension to the spinful case. Here we
give explicitly only the generalization of the expression
(\ref{rhoboson}) valid in the limit $L \rightarrow \infty$. One
now obtains
\begin{equation}
\rho (x) = \sqrt{\frac{2}{\pi}} \, \frac{\partial}{\partial x} \,
\vartheta_{\rho} (x)\, +\, \rho_{2k_F}(x)\, ,
\end{equation}
where
\begin{equation}
\rho_{2k_F}(x) = \frac{2k_F}{\pi}  \cos  \left[  2k_Fx+
\sqrt{2 \pi}  \vartheta_{\rho}(x) \right]  \cos \left[
\sqrt{2 \pi} \vartheta_{\sigma}(x) \right] \label{rhobosonspin}
\end{equation}
is the $2k_F$--contribution from the
interference between right-- and left--movers. Again we have suppressed
the ladder operators that need to be taken into account in general.

\section{Interaction, Voltage Bias, and Impurities in the Tomonaga--Luttinger
Model}

In the previous section we have investigated noninteracting
Fermions in a single channel quantum wire and seen that the
low energy physics of the Fermi gas can be described either 
in terms of occupation numbers of single electron states or
as excitations of Bosonic density waves. 
In higher dimensions noninteracting
quasiparticles, supplemented by an electroneutrality constraint,
give a rather accurate description also of the system in
presence of Coulomb interaction, provided some parameters are
replaced by effective parameters \cite{nozierespines}. 
It has been known since quite
some time that in one dimension the Fermi liquid description
breaks down. Remarkably, the interaction leads to rather moderate
modifications of the action in the Bose representation. The real
advantages of Bosonization will thus only emerge in this section
where we first present the real, interacting TL model and then
extend it to describe a quantum wire with a scatterer in presence
of an applied voltage.

\subsection{Electron-electron interaction and the
Tomonaga--Luttinger model} 

We now take the interaction into
account but restrict ourselves to spinless Fermions first. The
electronic charge density in a quantum wire is largely
compensated by a homogeneous positive background charge density.
Therefore, we now fix the Fermi wave number $k_F$ so that the
state $|N_+=0$, $N_-=0 \rangle_0$, where all single particle
states of $p$--movers with $pk \le k_F$ are occupied, is
electrically neutral. The density $\rho(x)$ introduced in
Eq.~(\ref{density}) multiplied by the electron change $e$ is then the
excess charge density in the wire, and the interaction may be
written
\begin{equation}
H_{{\rm int}} = \frac{1}{2} \, \int dx  \int dy \, \rho(x) \,
U(x-y) \, \rho(y) \, . \label{hint}
\end{equation}
The electron--electron interaction potential
\footnote{In the limit $L \rightarrow \infty$ the sum
$\frac{1}{L} \, \sum_q$ is replaced by $\frac{1}{2 \pi} \, \int
dq$\ .}
\begin{equation}
U(x) = \frac{1}{L} \, \sum_q \, U_q \, e^{iqx}
\end{equation}
has real Fourier components $U_q$ since $U(-x)=U(x)$.
Usually, the potential $U(x)$ deviates from a simple Coulomb
potential both at small and large distances. At small distances
of order the lateral dimensions of the quantum wire, the wave
function for the transversal motion of the strictly speaking
three--dimensional particles becomes relevant. When these
transversal components are integrated out, the resulting
effective potential $U(x)$ in the one--dimensional model remains
finite for $x \rightarrow 0$. On the other hand, at large
distances one has to take into account the effect of gate
electrodes and other nearby conductors that screen the long--range
part of the Coulomb interaction. The effective potential $U(x)$
then has a finite range $R$, which implies that the
logarithmic increase for small $q$ of the Fourier transformed 
Coulomb potential is
cut off at $q$--values of order $R^{-1}$. In particular, $U_{q=0}$
then remains finite, with the precise value depending on the
geometry of the problem. We assume that this externally screened
potential is still sufficiently long ranged so that $U_0 \gg
U_{2k_F}$. Then the $2k_F$--component of the electronic density
(\ref{rhoboson}) will give a negligible contribution when inserted
in the interaction (\ref{hint}), and we find for the Bosonized
interaction energy
\begin{equation}
H_{{\rm int}} = \frac{1}{2 \pi}  \int dx  \int dy \,
\frac{\partial \vartheta (x)}{\partial x} \, U(x-y) \,
\frac{\partial \vartheta (y)}{\partial y} \, . \label{intbose}
\end{equation}
When we restrict ourselves to low energy excitations with
wavelengths large compared to the range $R$ of the interaction
potential, we may replace $U(x)$ by a local interaction $U_0\,
\delta (x)$, and then obtain the interaction term of the TL model
\begin{equation}
H_{{\rm int}} = \frac{U_0}{2 \pi} \, \int dx  \left(
\frac{\partial \vartheta}{\partial x} \right)^2 \, . \label{inttl}
\end{equation}
We remark that a microscopic local interaction $\sim \delta (x)$
would of course have no effect on spinless Fermions as a
consequence of the anti--commu\-tation rules. Essentially, in
Eq.~(\ref{inttl}) one neglects the wave number dependence of $U_q$
for small $q$, but still $U_{2k_F} \ll U_0$. The interaction term
(\ref{inttl}) can readily be put in the action functional
(\ref{action}), which for the interacting model takes the form
\begin{equation}
S_{\rho} = \frac{\hbar}{2}  \int dt  \int dx \, \left[
\frac{1}{v_F}  \left( \frac{\partial \vartheta}{\partial t}
\right)^2 \, - v_F \left(  1 + \frac{U_0}{\pi \hbar v_F}
\right) \left( \frac{\partial \vartheta}{\partial x}
\right)^2 \right] \nonumber
\end{equation}
and thus remains of the form of the action of a harmonic string.
This clearly shows the great advantage of the Bose
representation: We still have a model of free Bosonic charge
density excitations, only the wave velocity is altered by the
interaction, while  in the Fermi representation the interaction
leads to quartic terms in the Fermi operators. 

It is customary to
introduce the coupling constant
\begin{equation}
g = \left( 1+ \frac{U_0}{\pi \hbar v_F} \right)^{- 1/2}
\label{gfaktor}
\end{equation}
and the charge density wave velocity
\begin{equation}
v = \frac{v_F}{g} \, . \label{vcharge}
\end{equation}
In terms of these quantities the action functional of the TL
model reads
\begin{equation}
S_{\rho} = \frac{\hbar}{2g}  \int dt  \int dx  \left[
\frac{1}{v}  \left( \frac{\partial \vartheta}{\partial t}
\right)^2 \, - v \left(  \frac{\partial \vartheta}{\partial x}
\right)^2 \right] \, . \label{actionvg}
\end{equation}
It would go beyond the scope of this article to demonstrate that
the action (\ref{actionvg}) indeed describes the low energy
properties of spinless one--dimensional Fermions correctly. Here
we refer to the literature \cite{voit,gogolin}. We would,
however, like to point out that the matter is in fact more
complex than what our plausible ``derivation'' of
Eq.~(\ref{actionvg}) might suggest. The Coulomb interaction is
strong and affects all states not only those near the two Fermi
points. Hence, it needs to be introduced in the underlying
physical model with a single energy dispersion curve
$\varepsilon_k$. Afterwards, one may integrate out states far
from the Fermi points until one reaches energy scales
sufficiently close to the Fermi energy to allow for a
linearization of the spectrum. At this point the Hamiltonian may
be re-written in terms of the right-- and left--movers of the TL
model, but the parameters of the model are then already
renormalized by the aforementioned elimination of high energy
excitations and additional interaction vertices are generated.
One can, however, conclude from a renormalization group study
\cite{solyom} that the action (\ref{actionvg}) is indeed a
low energy fixpoint of a (spin--polarized) quantum wire. From
these remarks it is clear that the parameter $U_0$ in
Eq.~(\ref{inttl}) does not necessarily coincide with the Fourier
coefficient at $q=0$ of the interaction potential. This latter
quantity can be considered an estimate of $U_0$ which becomes
more accurate for large electron densities at which the effect of
the Coulomb interaction is weaker. In the sequel we use $g$ and
$v$ as fundamental parameters of the model. For repulsive
interaction $U_0>0$ and thus $g<1$.
\subsection{Screening of external charges}
Let us assume that we perturb the quantum wire by an external
charge density
\begin{equation}
\rho_{{\rm ext}}(x,t) = e Q(x,t)\, , \nonumber
\end{equation}
which interacts with the electronic charge density $e \rho (x,t)$
via the same effective potential $U(x)$ introduced in the
previous section. We are interested in the long wavelength
response of the quantum wire and may thus disregard the
$2k_F$--component of the electronic charge density. With the local
approximation $U_0\, \delta (x)$, the action (\ref{actionvg}) is then
modified to read
\begin{equation}
S = S_{\rho}\,  -\, \frac{U_0}{\sqrt{\pi}}  \int dt  \int dx \,
Q(x,t) \, \frac{\partial}{\partial x} \, \vartheta (x,t) \, .
\label{actionnon}
\end{equation}
Since the action has only terms linear and quadratic in
$\vartheta$, the average electronic density $\langle \rho (x,t)
\rangle$ caused by $Q(x,t)$ can be determined from the phase
field $\overline{\vartheta} (x,t)$ minimizing the action
(\ref{actionnon}). The equation of motion
\begin{equation}
\left( \frac{\partial^2}{\partial t^2} \, - v^2
\frac{\partial^2}{\partial x^2}  \right) \,\overline{ \vartheta} (x,t) =
\frac{gv U_0}{\sqrt{\pi} \hbar} \, \frac{\partial}{\partial x} \,
Q(x,t)
\end{equation}
obeyed by the minimal action field is readily solved in terms of
the Fourier representation
\begin{equation}
\overline{\vartheta} (x,t) = \frac{1}{(2 \pi)^2}  \int dq
\int d \omega \, \tilde{\vartheta} (q, \omega) \, e^{iqx-i \omega
t} \, . \nonumber
\end{equation}
We find
\begin{equation}
\tilde{\vartheta}(q, \omega) = - \frac{g U_0}{\sqrt{\pi} \hbar}
\, \frac{ivq \, \tilde{Q}(q, \omega)}{\omega^2-v^2q^2} \, ,
\nonumber
\end{equation}
which yields for the electronic density $\langle \rho (x,t)
\rangle = \frac{1}{\sqrt{\pi}} \, \frac{\partial}{\partial x} \,
\overline{\vartheta} (x,t)$ in Fourier space
\begin{equation}
\langle \tilde{\rho} (q, \omega) \rangle = (1-g^2) \frac{v^2q^2
\tilde{Q}(q, \omega)}{\omega^2-v^2q^2} \, , \label{screenone}
\end{equation}
where we have expressed $U_0$ in terms of $g$ and $v$ by means of
Eqs.~(\ref{gfaktor}) and (\ref{vcharge}).

Now, the relation
between the external charge density $eQ(x,t)$ and the resulting
screening charge density $e \langle \rho (x,t) \rangle$ is
governed by the dielectric function
\begin{equation}
\varepsilon (q, \omega) = \frac{\tilde{Q}(q,
\omega)}{\tilde{Q}(q, \omega)+ \langle \tilde{\rho}(q, \omega)
\rangle} \, . \nonumber
\end{equation}
Combining this with Eq.~(\ref{screenone}) we find
\begin{equation}
\varepsilon (q, \omega) =
\frac{\omega^2-v^2q^2}{\omega^2-g^2v^2q^2} \, .
\end{equation}
In particular, in the static case $\omega =0$ we have
\begin{equation}
\varepsilon = \frac{1}{g^2} = 1 + \frac{U_0}{\pi \hbar v_F} \, ,
\end{equation}
which shows that the interaction parameter $g$ is directly
related to the dielectric constant of the quantum wire. In a
metallic system the dielectric function has a pole for $\omega
\rightarrow 0$, $q \rightarrow 0$ associated with the perfect
screening of static charges leading to electroneutrality.
However, in the TL model the long range part of the Coulomb
interaction is assumed to be screened by other conductors as
explained above. Then $U_0$ is finite and there is a finite
dielectric constant in the zero frequency and long wavelength
limits.

This is in accordance with the fact that
the total screening charge in units of $e$
\begin{equation}
Q_s = \int dx \, \langle \rho (x) \rangle = \langle
\tilde{\rho}(q=0) \rangle
\end{equation}
accumulated near a static impurity charge $Q$ follows from
Eq.~(\ref{screenone}) as \cite{eggergrabert2}
\begin{equation}
Q_s = - (1-g^2)Q \, .
\end{equation}
Hence a fraction $g^2Q$ of the external charge remains unscreened,
and the quantum wire is in general not electroneutral. As we will
discuss in greater detail in the next section, the charge $g^2Q$
is screened by the electrode responsible for the finite range of
the interaction. Formally, the limit of long range Coulomb
interaction corresponds to $g \rightarrow 0$ which implies
electroneutrality of the wire.

We mention that apart from the long
wavelength response of the quantum wire to an impurity charge
there is also a $2k_F$--response leading to Friedel oscillations
of the charge density. We will not discuss this here but refer to
the recent literature \cite{egger,saleur}.
\subsection{Electrostatics of a quantum wire}
As we have seen in the preceeding section,  the electrode
screening the long range part of the Coulomb interaction plays an
important role in the electrostatic response of a quantum wire to
external charges. We can visualize the TL model as a
one--dimensional quantum wire screened by a gate coupled to the
wire by a distributed capacitance as depicted in
Fig.~\ref{wiregate}. The interaction energy (\ref{inttl}) can
then be interpreted as the charging energy of the wire--gate
capacitance
\begin{equation}
H_{{\rm int}} = \frac{U_0}{2}  \int dx \, \rho(x)^2 \, =
\, \int dx \, \frac{e^2 \rho (x)^2}{2c_0} \, ,
\label{intcap}
\end{equation}
\begin{figure}
\hspace{1.4cm}
\epsfysize=4cm
\epsffile{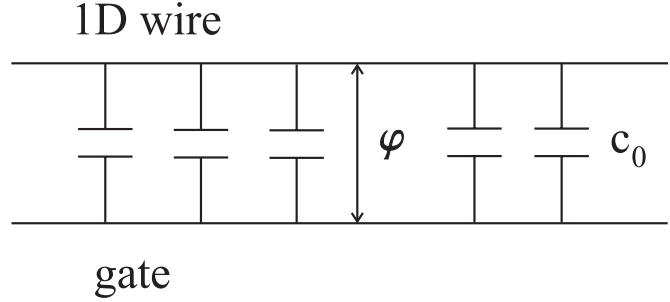}
\caption{Electrostatic model of a quantum wire
coupled by a capacitance per unit length $c_0$ to a gate. In a
nonequilibrium wire the band bottom is shifted by an electric
potential $\varphi$.}
\label{wiregate}
\end{figure}
where the capacitance per unit length $c_0$ is determined by
\begin{equation}
\frac{e^2}{c_0} = U_0 = \pi \hbar v_F \, \left( \, \frac{1}{g^2}
\, -1 \right) \, . \label{capacitance}
\end{equation}
An electronic charge density $e \rho (x)$ in the wire polarizes the
capacitance, and the resulting electric potential
\begin{equation}
\varphi (x) = \frac{e \rho (x)}{c_0} \label{electropot}
\end{equation}
shifts the band bottom of the quantum wire.
This is directly related to the
underscreening of external charges discussed previously as can be
seen from the following consideration. In a noninteracting system
an increase of the Fermi energy by $\Delta E$ shifts the wave
number of the Fermi points from $\pm k_F$ to $\pm (k_F+ \Delta
k)$ where $\Delta k = \Delta E/\hbar v_F$. There are $(L/2 \pi)
\Delta k$ single particle states in the wave number interval
$\Delta k$ and thus the density of $p$--movers increases by
\begin{equation}
\Delta \rho_p = \frac{\Delta k}{2 \pi} \, = \frac{\Delta E}{2 \pi
\hbar v_F} \, . \nonumber
\end{equation}
Here the factor $1/2 \pi \hbar v_F$ gives the density of states of
$p$--movers at the Fermi energy. Accordingly, the electronic
density increases by
\begin{equation}
\Delta \rho = \Delta \rho_+ + \Delta \rho_- = \frac{ \Delta E}{\pi \hbar
v_F} \, .\label{baredensity}
\end{equation}
On the other hand, in an interacting, electroneutral system the shift of the
Fermi energy by $\Delta E$ is accompanied by a shift of the band
bottom by the same amount and the electronic density remains
unchanged. In the TL model the situation is in between these
two extremal cases. The band bottom is shifted by $e
\varphi$ where $\varphi$ is the electric potential difference
between the wire and the gate electrode. Then, the change of the
electronic density is determined by
\begin{equation}
\Delta \rho = \frac{\Delta E - e \varphi}{\pi \hbar v_F} \,
\, . \nonumber
\end{equation}
In view of the relations (\ref{capacitance}) and (\ref{electropot}) we have
\begin{equation}
e \varphi = \frac{e^2}{c_0} \, \Delta \rho = \pi \hbar v_F
\left( \frac{1}{g^2} \, -1 \right) \Delta \rho \, ,\label{deltaphi}
\end{equation}
which gives
\begin{equation}
\Delta \rho = \frac{g^2 \Delta E }{\pi \hbar v_F}  \, . \label{deltarho}
\end{equation}
Hence, we see again that a fraction $g^2$ of the ``bare'' charge
density (\ref{baredensity}) caused by the shift of the Fermi level 
persists as true electronic charge density in the wire.

\subsection{Voltage bias and boundary conditions}

It is not difficult to generalize the preceeding considerations
to a nonequilibrium quantum wire in presence of an applied
voltage. Let us consider a quantum wire which is attached at the
ends to two-- or three--dimensional Fermi liquid reservoirs. We
assume that the contacts between the wire and the reservoirs are
adiabatic, which means that at the ends the quantum wire widens
sufficiently slowly to avoid any backscattering of outgoing
particles into the wire. This is the usual assumption underlying
Landauer's approach \cite{landauer} to the conductance of
mesoscopic wires. If the contacts are not adiabatic, there will
be an additional resistance depending on the precise realization
of the contacts and not only on intrinsic properties of the
quantum wire.

In equilibrium there is an equal amount of right--
and left--movers in the wire and the electrochemical potential is
constant. When we attach the wire to reservoirs, the influx of
right--movers at the left end of the wire will depend on the
electrochemical potential of the left electrode which we assume
to be $e U_L$ above the Fermi energy of the equilibrium quantum
wire. $U_L$ is then the voltage between the left reservoir and
the gate electrode screening the wire.
In the absence of interactions, the shift
of the Fermi energy by $eU_L$ would increase the density of
right--movers near the left end of the wire by
\begin{equation}
\rho^{{\rm bare}}_+ = \frac{eU_L}{2 \pi \hbar v_F} \label{barerho}
\end{equation}
as depicted schematically in Fig.~(\ref{voltage})a. Note that
below we will consider quantum wires with impurities. Then, the
reservoir determines the density of incoming particles only in
the clean section of the wire near the end, where the incoming
particles have not yet interacted with the impurities. The
density of outflowing particles, on the other hand, will be
affected by impurities.

In the presence of Coulomb interaction
the excess charge density caused by the reservoirs will charge
the distributed wire--gate capacitance leading to a shift of the
band bottom by
\begin{equation}
e \varphi = \frac{e^2}{c_0} \, \rho = \pi \hbar v_F  \left( 
\frac{1}{g^2} \, -1 \right)  \rho \ ,\label{shiftbot}
\end{equation}
where $e \rho$ is the true charge density in the wire that has to
be determined selfconsistently. The true density of right--movers
near the left end of the wire is then, cf.~Fig.~(\ref{voltage})b,
\begin{equation}
\rho_+ = \frac{e (U_L - \varphi)}{2 \pi \hbar v_F}\, ,
\label{righttrue}
\end{equation}
\begin{figure}
\hspace{1.5cm}
\epsfysize=8cm
\epsffile{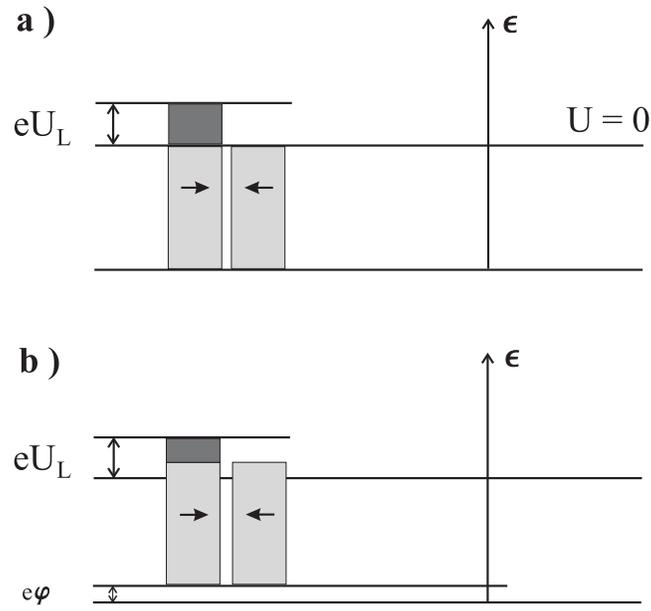}
\caption{Density of right--movers in a quantum
wire attached at the left end to a reservoir with voltage $U_L$
against the gate electrode. The equilibrium density is
represented as light-grey area and the nonequilibrium excess density
as dark-grey area. a) Sketch of the situation in the absence of
Coulomb effects and b) with interaction taking the shift $e
\varphi$ of the band bottom into account.}
\label{voltage}
\end{figure}
\noindent which means that the bare charge density (\ref{barerho})
is partially screened.

In terms of the Bosonic phase field
$\vartheta$, we have by virtue of the relations (\ref{ablthet})
\begin{equation}
\rho_+ = \frac{1}{2 \sqrt{\pi}} \left( \frac{\partial
\vartheta}{\partial x}  \, - \frac{1}{ v_F} \,
\frac{\partial \vartheta}{\partial t}\right) \, , \label{rightthet}
\end{equation}
where we have omitted the prime on $\vartheta$ as in all
equations following (\ref{ablthet}). We remark that in the
relation
\begin{equation}
\frac{1}{\sqrt{\pi}} \, \frac{\partial \vartheta}{\partial t} \,
= - v_F (\rho_+ - \rho_-) \nonumber
\end{equation}
the Fermi velocity $v_F$ is not altered by the Coulomb
interaction such as the velocity of charge density waves,  because
the microscopic expression of the particle current operator in
terms of Fermi operators is independent of the interaction in the
absence of a vector potential \cite{baym}. For the electric
potential $\varphi$ we obtain from Eqs.~(\ref{shiftbot}) and
(\ref{ablthet})
\begin{equation}
e \varphi = \sqrt{\pi} \hbar v_F \left( \frac{1}{g^2} \, -1
\right) \frac{\partial \vartheta}{\partial x} \, , \nonumber
\end{equation}
which can be combined with the relations (\ref{righttrue}) and
(\ref{rightthet}) to give for $x$ near the left end of the wire
\begin{equation}
\frac{1}{g^2} \, \frac{\partial \vartheta}{\partial x} \, -
\frac{1}{v_F} \, \frac{\partial \vartheta}{\partial t} \, =
\frac{e U_L}{\sqrt{\pi} \hbar v_F} \, . \nonumber
\end{equation}
Similar considerations hold for the density of left--movers near
the right end of the wire attached to an electrode at potential
$U_R$. Strictly speaking, all of the above considerations hold for
the average densities of right-- and left--movers. Hence, we find
that the coupling to reservoirs can be described in the Bosonized
TL model in terms of radiative boundary conditions for the phase
field \cite{eggergrabert}
\begin{eqnarray}
 \left( \frac{1}{g^2} \, \frac{\partial}{\partial x} \, -
\frac{1}{v_F} \, \frac{\partial}{\partial t}  \right) \,
\langle \vartheta (x,t) \rangle_{x=- \frac{L}{2}}& =& \frac{e
U_L}{\sqrt{\pi} \hbar v_F} \, , \nonumber \\
&& \label{boundary} \\
 \left( \frac{1}{g^2} \, \frac{\partial}{\partial x} \, +
\frac{1}{v_F} \, \frac{\partial}{\partial t}  \right) \, \langle
\vartheta (x,t) \rangle_{x = \frac{L}{2}}& =& \frac{e
U_R}{\sqrt{\pi} \hbar v_F} \, . \nonumber
\end{eqnarray}
If we impose these conditions at a point near the reservoirs,
they will be seen to be obeyed at any point in the impurity--free
clean sections at the ends of the wire.
\subsection{Impurity potential}
Nontrivial dc transport properties of the quantum
wire do not arise from the Coulomb interaction and the resulting
non--Fermi liquid behavior alone but only in connection with
impurities causing backscattering of electrons. A nonmagnetic
impurity at position $x_0$ couples to the charge density and gives
rise to an energy
\begin{equation}
H_W = \int dx \, W(x-x_0)\, \rho (x) \, , \label{wimp}
\end{equation}
where $W(x)$ is the impurity potential. When we want to add
this coupling term to the TL model, we have to note that the
perturbation affects all states, also those far from the Fermi
points, and we should again start from the underlying physical model
with a single branch. Applying essentially the same line of
reasoning used to include electron--electron interactions in the
TL model, we obtain for the case of an interaction with an
impurity at position $x_0$
\begin{equation}
H_W = \frac{W_0}{\sqrt{\pi}} \, \frac{\partial
\vartheta}{\partial x} \, \bigg|_{x=x_0} \, +
\frac{W_{2k_F}k_F}{\pi} \, \cos \left[ 2k_Fx_0+2 \sqrt{\pi}
\vartheta (x_0) \right] \, . \label{impbose}
\end{equation}
Of course, this form results when we insert the representation
(\ref{rhoboson}) of the electronic density into
Eq.~(\ref{wimp}) and then replace the Fourier coefficients $W_q$
of the impurity potential by constants for $q \approx 0$ and $|q|
\approx 2k_F$, respectively. Since an elimination of states far
from the Fermi points is necessary before we can rewrite the
Hamiltonian in terms of the Fermi operators $c^{}_{p,k} ,
c_{p,k}^{\dagger}$ of the TL model, the coefficients $W_0$ and
$W_{2k_F}$ must be interpreted as effective parameters that are
not directly related to Fourier coefficients of the microscopic
interaction potential. In addition, the elimination of states
will generate higher order processes involving several electrons
with momentum transfer $4k_F$, $6k_F$, and so on. It turns out
that the $2k_F$--processes dominate at low temperatures
\cite{kanefisher}.

Although an impurity is typically laterally
displaced from the center of the wire, the potential $W(x)$ will
essentially have the same behavior as the electron--electron
potential $U(x)$ discussed previously. Thus, from the bare
Fourier components we would conclude $W_{2k_F} \ll W_0$. But, as
we shall see, the forward scattering $W_0$, which scatters a
$p$--mover into another state of the same branch, has no effect
on the transport properties of the wire, while even a small
backscattering term $W_{2k_F}$ has a dramatic effect at
low energy scales \cite{kanefisher}.

To show that forward scattering is
unimportant, we write the TL Hamiltonian in presence of an
impurity in the form
\begin{eqnarray}
H & = & \frac{\hbar v_F}{2}  \int dx  \left[  \left( \,
\frac{\partial \phi}{\partial x}  \right)^2 \, + \left(
\frac{\partial \vartheta}{\partial x}  \right)^2 \, \right] \,
+ \, \frac{U_0}{2 \pi}  \int dx \, \left( \frac{\partial
\vartheta}{\partial x}  \right)^2 \nonumber \\
&& \label{hamtot} \\
&& + \ \frac{W_0}{\sqrt{\pi}} \, \frac{\partial
\vartheta}{\partial x} \, \bigg|_{x=x_0} \, +
\frac{W_{2k_F}k_F}{\pi} \, \cos \left[ 2k_F x_0 +2 \sqrt{\pi}
\vartheta (x_0) \right] \, , \nonumber
\end{eqnarray}
which includes the Hamiltonian (\ref{hamnull}) of the
noninteracting model, the interaction (\ref{inttl}), and the
impurity term (\ref{impbose}). Now, the unitary transformation
\begin{equation}
{\cal U} = \exp \left[ \, -i \sqrt{\pi} \int dx \, \alpha (x)
\, \phi (x) \right] \label{unitary}
\end{equation}
shifts the charge density, since
\begin{equation}
{\cal U} \,\frac{\partial \vartheta}{\partial x} \, {\cal U}^{-1} =
\frac{\partial \vartheta}{\partial x} \, + \sqrt{\pi} \, \alpha
(x) \, . \label{shiftrho}
\end{equation}
This is readily shown by considering the auxiliary function
\begin{equation}
F(s) = e^{isA} \, \frac{\partial \vartheta}{\partial x} \,
e^{-isA} \, , \nonumber
\end{equation}
where $A = \sqrt{\pi}  \int dx \, \alpha (x)  \phi (x)$ is
the exponent of ${\cal U}$. One then finds
\begin{equation}
\frac{\partial}{\partial s} \, F(s) = e^{isA} \, i \left[ A\,,
\, \partial \vartheta/ \partial x  \right]_- \, e^{-isA}
= \sqrt{\pi} \, \alpha (x) \, , \nonumber
\end{equation}
where the commutator is evaluated by means of
Eq.~(\ref{commphithet}). Since $F(0) = \partial \vartheta /
\partial x$, we find $F(1) = \partial \vartheta / \partial x +
\sqrt{\pi} \, \alpha (x)$ which is just the relation
(\ref{shiftrho}).

Based on the same commutator, one also finds
\begin{equation}
{\cal U} \, e^{2i \sqrt{\pi} \vartheta (x)} \, {\cal U}^{-1} =
e^{2i \sqrt{\pi} \vartheta (x) + i \eta (x)} \, , \label{exporho}
\end{equation}
where
\begin{equation}
\eta (x) = \pi  \int dy \ {\rm sign} (x-y) \, \alpha (y) \, .
\nonumber
\end{equation}
To see this, we write the left hand side of Eq.~(\ref{exporho})
as $e^{iA} \, e^{iB} \, e^{-iA}$ with the operator $A$ introduced
previously and $B = 2 \sqrt{\pi} \vartheta (x)$. Then, using
twice the relation (\ref{disent}), we find
\begin{equation}
e^{iA} \, e^{iB} \, e^{-iA} =  e^{iA} \, e^{i(B-A)} \, e^{-
\frac{1}{2}  [ A,B]_-}  =  e^{iB-[A,B]_-} \, , \nonumber
\end{equation}
which gives the transformation (\ref{exporho}).

With the help of the relations (\ref{shiftrho}) and
(\ref{exporho}), the Hamiltonian (\ref{hamtot}) is transformed into
\begin{eqnarray}
&& {\cal U}\, H\, {\cal U}^{-1} = \frac{\hbar v_F}{2}  \int dx
\left[ \left( \frac{\partial \phi}{\partial x}  \right)^2
\, + \, \frac{1}{g^2}  \left(  \frac{\partial \vartheta}{\partial
x}  \right)^2 \right] \nonumber \\
&& \nonumber \\
&& + \frac{W_0}{\sqrt{\pi}} \, \frac{\partial \vartheta}{\partial
x} \, \bigg|_{x=x_0} + \frac{\sqrt{\pi} \hbar v_F}{g^2} \int
dx \, \alpha (x) \, \frac{\partial \vartheta}{\partial x}
\label{htrans} \\
&& \nonumber \\
&& + \frac{W_{2k_F}k_F}{\pi} \, \cos  \left[ 2k_Fx_0 + 2
\sqrt{\pi} \vartheta (x_0) + \eta (x_0) \right] \, ,
\nonumber
\end{eqnarray}
where we have omitted terms that depend only on $\alpha (x)$ but
not on the phase fields. Further, we have introduced the
interaction parameter (\ref{gfaktor}). Now, with the choice
\begin{equation}
\alpha (x) = - \frac{g^2 W_0}{\pi \hbar v_F} \, \delta (x-x_0)
\end{equation}
the terms in the second line of Eq.~(\ref{htrans}) cancel, and we
are left with the Hamiltonian
\begin{eqnarray}
H^{\prime} & = & \frac{\hbar v_F}{2}  \int dx  \left[
\left( \frac{\partial \phi}{\partial x} \right)^2 +
\frac{1}{g^2}  \left(  \frac{\partial \vartheta}{\partial x}
 \right)^2 \right]    \label{hamprime}  \\
&& \nonumber \\
& + & \frac{W_{2k_F}k_F}{\pi}  \cos \left[  2k_Fx_0 + 2
\sqrt{\pi} \vartheta (x_0)  \right] \, ,\nonumber
\end{eqnarray}
which contains only a backscattering term. Note that for an
impurity with charge $eQ$ giving rise to the Coulomb
potential $W(x) = QU(x)$, the quantity
\begin{equation}
\alpha (x) = - \frac{g^2U_0}{\pi \hbar v_F} \, Q \, \delta (x-x_0) =
- (1-g^2) Q\, \delta (x-x_0) \nonumber
\end{equation}
is just the screening charge density (\ref{screenone}) caused by the
static impurity charge. The unitary transformation thus removes
the screening cloud, which is the main effect of the forward
scattering term. Below we will determine the
current--voltage relation of a quantum wire in presence of a
single impurity at position $x=0$. This study will be
based on the action
\begin{eqnarray}
S & = & \frac{\hbar}{2g}  \int dt  \int dx \, \left[
\frac{1}{v}  \left( \frac{\partial \vartheta}{\partial t}
\right)^2 \, - v \left( \frac{\partial \vartheta}{\partial x}
\right)^2 \right] \nonumber \\
&& \label{actionimp} \\
&-& \lambda \int dt \, \cos  \left[  2 \sqrt{\pi} \vartheta
(x=0,t) \right] \nonumber
\end{eqnarray}
associated with the Hamiltonian (\ref{hamprime}), where $\lambda$
characterizes the impurity strength.

\subsection{Interacting electrons with spin}

Here we briefly summarize the changes necessary to include the
electron spin. For noninteracting electrons we found that the
action of spinful electrons can be split into a charge and a
spin contribution. Since the Coulomb interaction couples only to
the charge density, we might argue that the charge part of the
action is modified by the interaction in the same way as for
spinless electrons while the spin part remains unchanged.
However, this would ignore the fact that the Coulomb interaction
must be introduced in the underlying physical model and the
transcription to TL Fermions can only be made after an
elimination of high energy excitations. This has consequences
also for the spin density waves, in particular, the spin wave
velocity $v_{\sigma}$ becomes smaller than the Fermi velocity
$v_F$ (\cite{hausler}). The low energy physics is then governed
by the action
\begin{eqnarray}
S & = & \frac{\hbar}{2g}  \int dt  \int dx \, \left[
\frac{1}{v}  \left(  \frac{\partial
\vartheta_{\rho}}{\partial t}  \right)^2 \, - v \left(
\frac{\partial \vartheta_{\rho}}{\partial x} \right)^2 \right]
\nonumber \\
&&  \\
&& + \frac{\hbar}{2}  \int dt  \int dx \, \left[
\frac{1}{v_{\sigma}}  \left( \frac{\partial
\vartheta_{\sigma}}{\partial t} \right)^2 \, - v_{\sigma}
\left(\frac{\partial
\vartheta_{\sigma}}{\partial x}\right)^2 \right] \, .
\nonumber
\end{eqnarray}
The TL model is thus characterized by three parameters $v$,
$v_{\sigma}$ and $g$. There is no coupling
constant $g_{\sigma}$ for the spin sector, which can be traced back to spin
rotation invariance \cite{voit,gogolin}. A detailed discussion of
these issues would go beyond the scope of this article and we
refer to the literature cited.  As in the noninteracting case, the
action splits into a charge and a spin part. However,  the
difference between the charge and spin wave velocities in an
interacting wire has important consequences and leads to the
notable phenomenon of spin--charge separation. When an electron
is injected into a wire it causes a charge and a spin pulse
propagating with different velocities.

In case an impurity is added at position $x=0$ to the model
we see from the $2k_F$--part of the electron density
 in the spinful case (\ref{rhobosonspin}) that the important
backscattering term now has the form
\begin{equation}
S_{\lambda} = - \lambda \int dt \, \cos  \left[ \sqrt{2
\pi} \vartheta_{\rho}(x=0,t) \right] \, \cos  \left[ \sqrt{2
\pi} \vartheta_{\sigma} (x=0,t) \right] \, . \nonumber
\end{equation}
Hence, the impurity couples the charge and spin sectors making
the theory of dirty quantum wires in the absence of a
spin--polarizing magnetic field more involved.

\section{Current--voltage relation of a quantum wire}

In this section we apply the theory developed so far and
determine the current in a quantum wire with an impurity as a
function of the applied voltage and the temperature. Rather than
giving an overview of the results available in the literature, we
treat a special case in some detail to illustrate how the
formalism explained in the previous sections can be employed to
obtain concrete results. With this background readers should then
be prepared to embark on reading the recent original literature
on the subject.
\subsection{Particular solution and four--terminal voltage}
We study a single channel quantum wire with an impurity at
position $x=0$. The conductor is attached to reservoirs at
voltages $U_L$ and $U_R$ relative to the gate electrode screening
the wire as sketched in Fig.~(\ref{wire}). The applied voltage
\begin{equation}
U=U_L-U_R  \nonumber
\end{equation}
will then drive a current $I$ through the wire. For simplicity,
we shall restrict ourselves to the case of spinless electrons. We
can then base the consideration on the action functional
(\ref{actionimp}) and the boundary conditions (\ref{boundary}).
\begin{figure}
\hspace{0.7cm}
\epsfysize=4cm
\epsffile{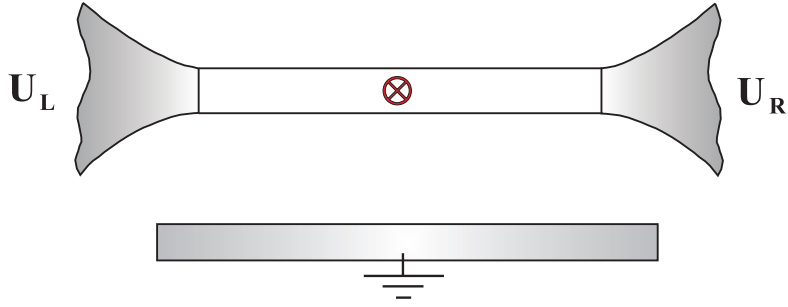}
\caption{Gated quantum wire adiabatically connected
to reservoirs at voltages $U_L$ and $U_R$. A backscatterer is
located in the center of the wire.}
\label{wire}
\end{figure}

As discussed previously, the phase field $\vartheta (x,t)$
describes also the mean particle densities (\ref{rhop}) due to a
shift introduced in Eq.~(\ref{thetashift}). To take care of the
boundary conditions, we look for a particular field $\theta
(x,t)$ that satisfies the wave equation in the clean parts of the
wire and the boundary conditions. This solution is of the form of
the shift in Eq.~(\ref{thetashift})
\begin{equation}
\theta (x,t) = \frac{g^2e}{2\sqrt{\pi}\hbar v_F} \left[
(U_L+U_R)x-V|x| \right] \, - \,
 \frac{e}{2\sqrt{\pi}\hbar}(U_L-U_R-V)t
 \, , \label{part}
\end{equation}
where $V$ is an arbitrary parameter. We now split the phase
field $\vartheta (x,t)$ into
\begin{equation}
\vartheta (x,t) = \theta (x,t) + \varphi (x,t) \, . \label{partphi}
\end{equation}
The deviation $\varphi (x,t)$ from the particular solution
(\ref{part}) will then obey equilibrium boundary conditions.
Further, we fix the parameter $V$ by the requirement that in the
steady nonequilibrium state
\begin{equation}
\frac{\partial}{\partial t} \, \langle \varphi (x,t) \rangle = 0
\, . \label{condition}
\end{equation}
Now, the average current $I$ may be written as
\begin{equation}
I = ev_F \langle \rho_+ - \rho_- \rangle = - \frac{e}{\sqrt{\pi}}
\, \frac{\partial}{\partial t} \, \langle \vartheta \rangle \, ,
\nonumber
\end{equation}
where we have used Eq.~(\ref{ablthet}) to obtain the second
equality. Hence, the condition (\ref{condition}) means that
the average current
is determined solely by the particular solution (\ref{part}),
and we have
\begin{equation}
I = \frac{e^2}{h}(U-V)\, . \label{ivu}
\end{equation}
Since in a steady state the current $I$ is independent of $x$, we
may impose the condition (\ref{condition}) fixing the voltage $V$
at any point $x$.

The particular solution (\ref{part}) not only
determines the average current but also the average charge
density in the nonequilibrium quantum wire. Since $\langle
\varphi \rangle$ obeys equilibrium boundary conditions as well as the
condition (\ref{condition}), it does not contribute to the
average density (\ref{rhoboson})
\begin{equation}
\langle \rho \rangle = \frac{1}{\sqrt{\pi}} \,
\frac{\partial}{\partial x} \, \langle \vartheta \rangle \, ,
\nonumber
\end{equation}
where we have omitted the $2k_F$-component which gives an
additional oscillatory contribution near the impurity. This
Friedel oscillation component is not seen in a density smoothed
over length scales of order $\lambda_F = 2 \pi/k_F$. From the
particular solution (\ref{part}) we obtain for the average density
\begin{equation}
\langle \rho \rangle = \frac{g^2e(U_L+U_R)}{2 \pi \hbar v_F} \, -
\, \frac{g^2eV}{2 \pi \hbar v_F} \, {\rm sign} (x) \, . \nonumber
\end{equation}
The first term just describes the change (\ref{deltarho})
of the average electronic density as a consequence of the average
shift $e(U_L+U_R)/2$ of the Fermi energy. This term is absent if
the voltage $U$ is applied asymmetrically, i.e., for $U_L=-U_R=U/2$. The
second term gives an asymmetric component of the charge density in
presence of an impurity. The density drop
\begin{equation}
\Delta \rho = \frac{g^2eV}{\pi \hbar v_F} \nonumber
\end{equation}
across the impurity site is associated with a difference
\begin{equation}
\Delta \mu = g^2 eV \nonumber
\end{equation}
of the effective chemical potential on both sides of the
impurity. Furthermore, the drop of the
charge density across the impurity site is also associated with a
change of the electric potential (\ref{shiftbot}) by
\begin{equation}
\Delta \varphi = \frac{\pi \hbar v_F}{e}  \left(
\frac{1}{g^2} \, - 1 \right)  \Delta \rho = (1-g^2) \, V \, .
\nonumber
\end{equation}
In a hypothetical ideal measurement of the voltage drop across
the impurity site one would observe the difference of the
electrochemical potential \cite{landauer}
\begin{equation}
\Delta \varphi + \frac{1}{e} \, \Delta \mu = V \, . \nonumber
\end{equation}
Hence, the parameter $V$ introduced above coincides with the
average four-terminal voltage $V$, which is the part of the
applied voltage $U$ dropping across the 
scatterer.\footnote{The readers should be aware of the fact that 
in some early treatments of transport properties of the TL model
the discrimination between $U$ and $V$ was not made.}

In view of Eq.~(\ref{ivu}) the determination
of the current--voltage relation corresponds to a calculation of
the four--terminal voltage. Two limiting cases are evident from
physical grounds. In the absence of a backscatterer $(\lambda
\rightarrow 0)$ we have $V=0$ and obtain from Eq.~(\ref{ivu})
\begin{equation}
I=G_0 \, U \, , \nonumber
\end{equation}
where $G_0 = e^2/h$ is the conductance of a clean wire. This is
the same result as obtained previously for noninteracting
electrons. Hence, for a clean quantum wire with adiabatic
contacts to the reservoirs the interaction has no effect on the
conductance
\cite{maslovstone,ponamarenko,safischulz,eggergrabert}. On the
other hand, for a very strong backscatterer $(\lambda \rightarrow
\infty)$ we have $V=U$ and the current $I$ vanishes. In the
remainder we shall discuss how we get from one limit to the other.
\subsection{Path integral on the Keldysh contour}
To treat the nonequilibrium quantum wire with the action
functional (\ref{actionimp}) quantum mechanically, we have to
evaluate a Feynman path integral on the Keldysh contour. For an
introduction to the Keldysh technique we refer to the review article
\cite{keldysh}, however, the basic idea can be understood in the
following way. Assume that at time $t_0$ the system is described
by the density matrix $W(t_0)$ and let $H$ be the Hamiltonian
including the coupling to the reservoirs. The density matrix at a
later time $t_f$ is then given by
\begin{equation}
W(t_f) = e^{- \frac{i}{\hbar} H(t_f-t_0))} \, W(t_0)\,
e^{\frac{i}{\hbar} H(t_f-t_0)} \, . \label{dmatrix}
\end{equation}
Each of the two time evolution operators $e^{\pm \frac{i}{\hbar}
H(t_f-t_0)}$ may be written as a Feynman path integral. Since
we are interested in steady state properties independent of the
initial state $W(t_0)$, we take the limit $t_0 \rightarrow -
\infty$. The trace over such a time propagated operator leads to
a path integral of the form
\begin{equation}
Z_0 = \int {\cal D} [\vartheta] \, e^{\frac{i}{\hbar}
S[\vartheta]} \, , \label{pathint}
\end{equation}
where $S[\vartheta]$ is the action functional (\ref{actionimp})
with the time integration $\int dt$ running along the Keldysh
contour depicted  in Fig.~(\ref{keldysh}).
\begin{figure}
\hspace{0.3cm}
\epsfysize=1.4cm
\epsffile{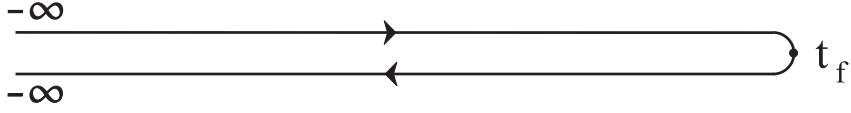}
\caption{The Keldysh contour runs from $- \infty$
to $t_f$ along the real time axis and back to $- \infty$. The two
branches of the contour correspond to the forward and backward
propagators in Eq.~(\ref{dmatrix}).}
\label{keldysh}
\end{figure}
In order to employ this path integral for the calculation of
expectation values, we first decompose the phase field according
to Eq.~(\ref{partphi}). The action (\ref{actionimp}) then takes
the form
\begin{eqnarray}
S & = & \frac{\hbar}{2g} \int dt  \int dx \, \left[
\frac{1}{v} \left( \frac{\partial \varphi}{\partial t}
\right)^2 \, - v \left( \frac{\partial \varphi}{\partial x}
\right)^2 \, \right] \nonumber \\
&& - \frac{eV}{\sqrt{\pi}}  \int dt \, \varphi (0,t)
\label{actiontrans} \\
&& - \lambda  \int dt \, \cos \left[  2 \sqrt{\pi} \, \varphi
(0,t) \, - \frac{e}{\hbar}  (U-V)t  \right] \, , \nonumber
\end{eqnarray}
where we have omitted terms independent of $\varphi (x,t)$.
Further, we have taken into account that
\begin{equation}
\int dt \, \frac{\partial \varphi}{\partial t} \, = 0
\end{equation}
for a time integral along the Keldysh contour and
\begin{equation}
\int dx \, {\rm sign} (x) \frac{\partial}{\partial x} \, \varphi
(x,t) = -2 \varphi (0,t) \, . \nonumber
\end{equation}
For the phase field $\varphi (x,t)$, which obeys
equilibrium boundary conditions, the nonequilibrium situation
becomes apparent in two modifications of the action
(\ref{actiontrans}). The term in the second line comes from the
voltage drop across the impurity site. It describes the
potential energy
\begin{equation}
\int dx \, \left[  - \frac{V}{2} \, {\rm sign} (x) \right] \,
\frac{e}{\sqrt{\pi}} \, \frac{\partial \varphi (x,t)}{\partial x}
= \frac{eV}{\sqrt{\pi}} \, \varphi (0,t)
\end{equation}
of a charge density fluctuation $(e/\sqrt{\pi}) \, \frac{\partial
\varphi}{\partial x}$ in presence of an electrochemical potential
$- (V/2)\, {\rm sign} (x)$.  The remainder $U-V$ of the
applied voltage $U$ shows up in the third line of Eq.~(\ref{actiontrans})
as a Josephson--type phase shift in
the pinning potential caused by the impurity. 

Instead of the path integral (\ref{pathint}) we now study the functional
\begin{equation}
Z [ \eta ] = \int {\cal D} [ \varphi ] \, e^{\frac{i}{\hbar}
S[\varphi] + i \sqrt{\pi}  \int dt \, \eta (t) \,
\frac{\partial}{\partial t}  \varphi (0,t)} \, ,
\label{genfunct}
\end{equation}
where $S[ \varphi ]$ is the action (\ref{actiontrans}), and where
we have introduced an auxiliary field $\eta (t)$ on the Keldysh
contour. $Z[\eta]$ is a generating functional for expectation
values of $\frac{\partial}{\partial t} \, \varphi (0,t)$. In
particular, the condition (\ref{condition}) is now equivalent to
\begin{equation}
\frac{\delta Z[\eta]}{\delta \eta (t)} \, \bigg|_{\eta=0} = 0
\, , \label{conditionz}
\end{equation}
which needs to be evaluated to determine the four--terminal
voltage $V$.

\subsection{Coulomb gas representation}

In the sequel we outline one of the methods available to determine
the current--voltage relation. While for the one impurity
problem considered here, an approach based on the
thermodynamic Bethe ansatz is most powerful \cite{eggersaleur},
we present here a technique which remains useful also for multi
impurity problems. First, we split the Keldysh contour explicitly
into the two branches and denote the phase
field on the branch from $- \infty$ to $t_f$ by $\varphi (x,t)$
and the field on the branch from $t_f$ back to $- \infty$ by
$\varphi^{\prime} (x,t)$. Further, we split the action 
(\ref{actiontrans}) into two terms
\begin{eqnarray}
S_0\! & = &\! \frac{\hbar}{2g} \int^{t_f}_{- \infty}  dt  \int
dx  \left[  \frac{1}{v}  \left( \frac{\partial
\varphi}{\partial t}  \right)^2 \!\! - v \left(  \frac{\partial
\varphi}{\partial x}  \right)^2 \!\! - \frac{1}{v}  \left(
\frac{\partial \varphi^{\prime}}{\partial t}  \right)^2 \!\! + v
\left(  \frac{\partial \varphi^{\prime}}{\partial x}
\right)^2  \right] \nonumber \\
&&\label{esnull} \\
&& - \frac{eV}{\sqrt{\pi}} \ \int^{t_f}_{- \infty} \, dt \,
\left[  \varphi (0,t) - \varphi^{\prime}(0,t) \right]
\nonumber
\end{eqnarray}
and
\begin{eqnarray}
S_{\lambda} = - \lambda  \int^{t_f}_{- \infty} & dt & \left\{
\cos  \left[ 2 \sqrt{\pi}  \varphi (0,t) -
\frac{e}{\hbar}  (U-V)t \right] \right.  \nonumber\\
 && -\left.  \cos \left[ 2 \sqrt{\pi} \varphi^{\prime}(0,t) -
\frac{e}{\hbar}  (U-V)t \right] \right\} \, . \label{eslambda}
\end{eqnarray}
With the help of the trigonometric relation
\begin{equation}
\cos \alpha - \cos \beta = -2 \sin \frac{\alpha + \beta}{2} \,
\sin \frac{\alpha - \beta}{2} \label{trigo}
\end{equation}
the action $S_{\lambda}$ may be written as
\begin{equation}
S_{\lambda} = 2 \lambda \, \int^{t_f}_{- \infty} \, dt \, \cos
A(t) \, \sin B(t) \, , \label{escossin}
\end{equation}
where
\begin{eqnarray}
A(t) & = & \sqrt{\pi} \, [ \, \varphi (0,t) + \varphi^{\prime}
(0,t)] \, - \frac{e}{\hbar} \, (U-V)t + \delta \nonumber \\
&& \label{aundb} \\
B(t) & = & \sqrt{\pi} \, [ \, \varphi (0,t) -
\varphi^{\prime}(0,t) ] \, . \nonumber
\end{eqnarray}
From Eqs.~(\ref{eslambda}) and (\ref{trigo})  
we obtain $\delta = - \frac{\pi}{2}$, however,
the precise value of this phase must be irrelevant, since we can
always add a constant phase to the particular solution
(\ref{part}), e.g., by replacing $t$ by $t-t_0$. Such an additional
phase of $\theta (x,t)$ leads to a shift of $\delta$.

Using Eq.~(\ref{escossin})
we find by expanding in powers of $\lambda$
\begin{equation}
e^{\frac{i}{\hbar}  S_{\lambda}} \, = 1+ \sum^{\infty}_{n=1} \,
\left( \frac{2i \lambda}{\hbar} \right)^n  \int  {\cal
D}_nt \ \prod^n_{j=1} \, \cos A(t_j) \, \sin B (t_j) \, , \nonumber
\end{equation}
where we have introduced the abbreviation
\begin{equation}
\int  {\cal D}_nt = \int^{t_f}_{- \infty} \, dt_n \,
\int^{t_n}_{- \infty} \, dt_{n-1} \, \ldots \, \int^{t_2}_{-
\infty} \, dt_1 \ . \nonumber
\end{equation}
Next, we write the trigonometric functions as
\begin{eqnarray}
\cos A(t_j) & = & \frac{1}{2} \, \sum_{u_j = \pm} \,
e^{iu_jA(t_j)} \, ,
\nonumber \\
&&\label{sumuv} \\
\sin B(t_j) & = & \frac{1}{2i} \, \sum_{v_j= \pm} \, v_j \,
e^{iv_jB(t_j)} \, , \nonumber
\end{eqnarray}
which gives
\begin{equation}
e^{\frac{i}{\hbar}  S_{\lambda}} = 1+ \sum^{\infty}_{n=1} \,
\sum_{\{u_j,v_j \}}  \left(  \prod^n_{j=1}  \frac{\lambda
v_j}{2 \hbar} \right) \, \int {\cal D}_nt \
e^{\frac{i}{\hbar}  S_n} \, , \nonumber
\end{equation}
where
\begin{equation}
\frac{1}{\hbar}  S_n = \sum^n_{j=1} \, \left[  u_j  A(t_j)
+ v_j B(t_j)  \right]  \nonumber
\end{equation}
is linear in the phase fields $\varphi$, $\varphi^{\prime}$ by
virtue of Eq.~(\ref{aundb}). The benefit of this expansion is that now the
path integral (\ref{genfunct}) is Gaussian order by order.
Therefore, we can integrate out the $\varphi$ and
$\varphi^{\prime}$ fields. Essentially, the calculation of these
path integrals amounts to an explicit computation of the fields
minimizing the action.

On a formal level we may consider the
auxiliary variables $u_j$, $v_j$ as charges on the time axis that
are coupled to a harmonic string described by the phase fields.
The integration over the fields $\varphi$, $\varphi^{\prime}$
then corresponds to the elimination of a harmonic bath in the
context of dissipative quantum mechanics. This problem is well
studied in the literature \cite{feynmanvernon, caldeiraleggett,
report}. It would go beyond the scope of this article to present
these methods explicitly here. Once the Gaussian integrals over
the phase fields are carried out, the sum over the charges
$v_j$ can be done straightforwardly, and the generating
functional (\ref{genfunct}) is obtained in the form
\begin{eqnarray}
Z[\eta]& = & \exp  \left\{  - \int dt  \int^t dt^{\prime} \,
\eta (t) \, \ddot{C} (t-t^{\prime}) \, \eta (t^{\prime}) -ig\frac{e}{\hbar}V 
\int dt \, \eta (t) \right\} \nonumber \\
&& \\
&& \times \, \left(  1+ \sum^{\infty}_{m=1}  Z_m[\eta] 
\right) \, , \label{series} \nonumber
\end{eqnarray}
where
\begin{eqnarray}
 Z_m[\eta]& =& \left(\frac{i \lambda}{\hbar}\right)^{2m} \int
{\cal D}_{2m} t \,
\sum_{\{ u_j \}^{\prime}}  \exp\left(\sum^{2m}_{j>k=1}  u_ju_k  C(t_j-t_k)
\right. \nonumber \\
&& \left. + 
\sum^{2m}_{j=1}  u_j \left[ \int dt\, \eta (t) 
\dot{C}(t-t_j) -i\frac{e}{\hbar}  (U-V+gV)t_j  \right]\right) \nonumber \\
&& \nonumber \\
&& \times \sin\left[\pi g \eta(t_{2m})\right]
\prod^{2m-1}_{j=1} \, \sin \left( \pi g\left[ \eta (t_j) +
\sum_{k=j+1}^{2m} u_k\right] \right)  \label{zetn} 
\end{eqnarray}
Here the sine functions arise from the sum over the charges $v_j$
by means of Eq.~(\ref{sumuv}). The function
\begin{equation}
C(t) = 2g  \ln  \left[ \frac{\beta\Delta}{\pi } \, \sinh 
\left(  \frac{\pi |t|}{\hbar\beta}  \right) \right] \label{cint}
\end{equation}
describes an effective interaction between the charges $u_j$ whereby
$\beta = 1/k_BT$ is the inverse temperature of the quantum
wire and $\Delta$ is the cutoff energy of the harmonic string. 
The temperature emerges from the asymptotic conditions on the
fluctuations of the phase field for $t \rightarrow -\infty$.
To simplify
notation we have chosen the
same auxiliary field $\eta (t)$  on both branches of
the Keldysh contour so that the second term in the exponent of
(\ref{genfunct}) reads $\sqrt{\pi} \, \int^{t_f}_{-\infty} dt \,
\eta (t) \, \frac{\partial}{\partial t} \, [\varphi (0,t) +
\varphi^{\prime}(0,t)]$. Furthermore, we have taken the limit
$t_f \rightarrow \infty$. 

In the series (\ref{series}) only even
terms in $\lambda$ survive, because $C(t)$ is a long range
interaction which suppresses all terms that do not satisfy the
charge neutrality condition
\begin{equation}
\sum^n_{j=1} \, u_j = 0 \, . \label{neutral}
\end{equation}
Since the charges $u_j = \pm 1$, this condition can only hold for
$n=2m$. The constraint (\ref{neutral}) is indicated as a prime at
the sum over the $u_j$ in Eq.~(\ref{zetn}). Because of this condition
the phase $\delta$ in Eq.~(\ref{aundb}) drops out. The representation
(\ref{series}) of the generating functional is known as the
Coulomb gas representation of the problem, since some terms allude 
to the partition function of one--dimensional charges interacting
with the ``Coulomb potential'' (\ref{cint}). Note that in view
of the factors $\sin[\pi g \eta(t_{2m})]$ the generating
functional (\ref{series}) obeys the normalization $Z[\eta=0] = 0$.

The four--terminal
voltage $V$ can now be determined from the condition
(\ref{conditionz}). A nonvanishing contribution of $Z_m[\eta]$
only arises if the variational derivative acts upon 
$\sin[\pi g \eta(t_{2m})]$. Introducing the time--difference variables
\begin{equation}
\tau_m = t_{m+1} - t_m \nonumber
\end{equation}
we find
\begin{equation}
\frac{e}{\hbar} V= K\left(\frac{e}{\hbar}[U-V+gV]\right) \, , 
\label{vundk}
\end{equation}
where
\begin{eqnarray}
K(\Omega)& =& \pi \, {\rm Im} \, \sum^{\infty}_{m=1} \, (-1)^{m} \,
\left(\frac{\lambda}{\hbar}\right)^{2m} \, 
\int^{\infty}_0 d\tau_1\ \ldots \int^{\infty}_0
d\tau_{2m-1}  \label{ksum}\\
&&\nonumber \\
&& \times \sum_{\{u_j\}^{\prime}} \, \exp\left[\sum^{2m}_{j>k=1} 
u_ju_k  C\left(\sum^{j-1}_{l=k} \tau_l\right)\right] 
\prod^{2m-1}_{j=1} \, \sin(\pi g p_j) \, e^{-ip_j\Omega
\tau_j} \, . \nonumber
\end{eqnarray}
Here
\begin{equation}
p_j = \sum^{2m}_{k=j+1} \, u_k = - \sum^j_{k=1} \, u_k \nonumber
\end{equation}
is the accumulated charge in the time interval $t>t_j$.
Sums of the form (\ref{ksum}) with an
interaction (\ref{cint}) are again familiar from dissipative
quantum mechanics \cite{weiss}

\subsection{Exact solution for $g= \frac{1}{2}$}

The explicit evaluation of the sum (\ref{ksum}) is greatly
simplified in the special case $g= \frac{1}{2}$, since the
sine--functions then suppress many terms. The so--called
``collapsed blip approximation'' \cite{weiss} of the Coulomb gas
problem becomes exact for $g= \frac{1}{2}$ and $K(\Omega)$ can easily
be calculated explicitly. One finds
\begin{equation}
K(\Omega) = 2 \frac{\lambda_B}{\hbar} \ {\rm Im }\, \psi \left( 
\frac{1}{2} \, + \frac{\beta \left[{\lambda_B} +i\hbar\Omega\right] }{2 \pi} 
\right) \, , \nonumber
\end{equation}
where $\psi (z)$ is the digamma function and
\begin{equation}
{\lambda_B} = \pi \frac{\lambda^2}{\Delta} \nonumber
\end{equation}
is a renormalized energy scale of the impurity. The
four--terminal voltage $V$ then follows from Eq.~(\ref{vundk}), which
for $g= \frac{1}{2}$ reads
\begin{equation}
\frac{e}{\hbar}V=K \left(\frac{e}{\hbar}[ U- V/2]\right) \, . \nonumber
\end{equation}
It can be easily seen that the solution is of the form
\begin{equation}
\frac{eV}{{\lambda_B}} = f \left( \frac{eU}{{\lambda_B}}
 \, , \,
\frac{k_BT}{{\lambda_B}} \right) \, . \nonumber
\end{equation}
This shows that the
solution exhibits scaling. When the energies $eU$ and
$k_BT$ are measured in units of the renormalized  impurity
energy scale ${\lambda_B}$, we find the same four--terminal
voltage (in units of ${\lambda_B}$) for any impurity and
hence also the same form of the current--voltage relation
(\ref{ivu}).

In Fig.~(\ref{ivcurve}) we depict the
current--voltage curve for various temperatures. We see that the
interaction has a dramatic effect on the
$I$--$V$--characteristics in presence of a scatterer. While in
the noninteracting model the conductance is essentially
independent of $k_BT$ and $eU$ for small energy scales, there is a
\begin{figure}
\hspace{0.5cm}
\epsfysize=8cm
\epsffile{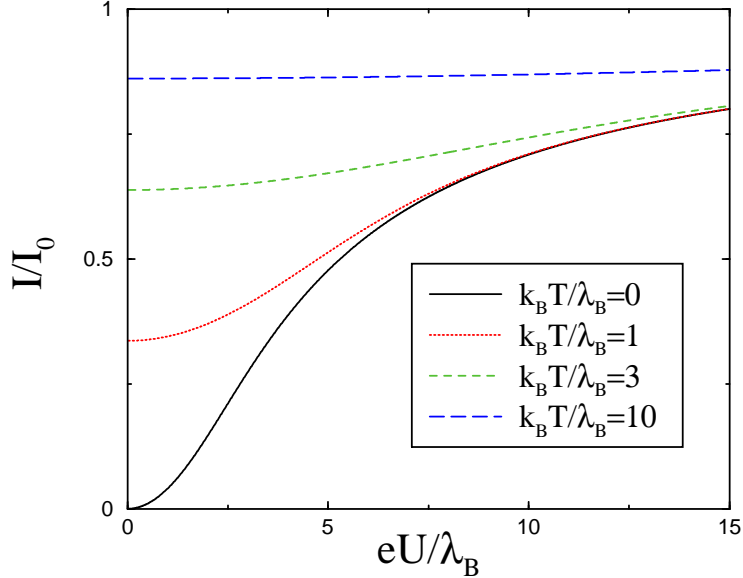}
\caption{Current--voltage characteristics for
several temperatures $T$. The current is normalized to $I_0 =
(e^2/h)U$.}
\label{ivcurve}
\end{figure}
strong suppression of the conductance in the interacting system
for energies below $\lambda_B$. At
zero temperature
\begin{equation}
\frac{I}{I_0} \, = \, \frac{1}{24} \left(
\frac{eU}{{\lambda_B}}  \right)^2 \ {\rm for} \ eU \ll
{\lambda_B} \nonumber
\end{equation}
and
\begin{equation}
\frac{I}{I_0} \, = \, 1- \pi \frac{{\lambda_B}}{eU} \ \ \ \ {\rm
for} \ eU \gg {\lambda_B} \, , \nonumber
\end{equation}
where $I_0 = (e^2/ h)U$. Hence, the differential conductance
shows a perfect zero bias anomaly and vanishes $\sim U^2$ for 
$U \rightarrow 0$. On the other hand, the conductance approaches the
value $e^2/h$ of a clean wire for large voltages. In practice,
this latter limit will not be reached, since for large $U$ the
low energy description in terms of the TL model breaks down.
On the other hand for $T>0$ the linear conductance for $U
\rightarrow 0$ reads
\begin{equation}
G(T) = \frac{e^2}{h} \, \frac{\Theta - \psi^{\prime}(\frac{1}{2}
\, + \frac{1}{\Theta})}{\Theta + \psi^{\prime}(\frac{1}{2} \, +
\frac{1}{\Theta})} \, , \nonumber
\end{equation}
where $\Theta=2 \pi k_BT/ {\lambda_B}$. Hence, the conductance
vanishes $\sim T^2$ for $k_BT \ll {\lambda_B}$. We
mention that the exact solution for $g = \frac{1}{2}$ \, was
first obtained by re--Fermionization techniques
\cite{eggergrabert}.
\subsection{Concluding remarks}
The explicit results for the current--voltage characteristics
derived in the preceeding section show that
for a spin--polarized quantum wire the crossover from
the weak impurity problem at higher temperatures or larger applied
voltages to the strong coupling problem at low temperatures and small
voltages can be solved exactly. The scaling function can also be 
obtained for arbitrary $g<1$ \cite{eggersaleur}. 
Mostly, one has to be content with more limited results.
In the spinful case the current is known only asymptotically 
for large and small energy scales \cite{kanefisher}. Several
topics are still under investigation. For instance, 
in the presence of many impurities additional features 
such as resonant tunneling may arise. 
Another area of active research is the current
noise in one--dimensional conductors.

We have focussed here on the physics of screened single
channel quantum wires at low energy scales where the finite
range of the interaction is unimportant. Many results can in 
fact be extended to finite range interactions. However, in
the absence of a gate, the long range nature of the
Coulomb interaction requires special attention \cite{schulz}. 
There are many similarities between the transport properties 
of quantum wires and edge currents in fractional quantum Hall 
bars \cite{chang}. Since in these latter
devices right-- and left--moving branches are spatially
separated, they are described by a chiral TL model \cite{wen,kane,fendley}
with somewhat different transport properties.
The methods presented here are also useful to model the
low energy electronic properties of carbon nanotubes
\cite{eggergog,balents}. Again, due to features of the
bandstructure, some differences arise leading to four channels
where one channel is charged as in a semiconductor quantum wire. 
We hope that this article will facilitate the study of these and
many other exciting new results on transport properties of
one--dimensional Fermionic systems.

\subsection*{Acknowledgments}
The author wishes to thank Reinhold Egger for an enjoyable 
collaboration on the subject of this article in
recent years. The manuscript has benefitted from a
critical reading by Wolfgang H{\"a}usler, J{\"o}rg Rollb{\"u}hler, 
and Bj{\"o}rn Trauzettel.
Financial support was provided by the Deutsche Forschungsgemeinschaft (Bonn).

\end{document}